\documentclass[12pt]{article}
\usepackage{graphicx,amsmath,amsthm,amscd,amsfonts}
\usepackage{axodraw}

\usepackage{amssymb}
\input xy
\xyoption{all}

\theoremstyle{definition}

\newcommand\cE{\mathcal{E}}

\begin{document}

\hfill 

\vspace{1.0in}

\begin{center}

{\large\bf Physical aspects of quantum sheaf cohomology}

{\large\bf for deformations of tangent bundles of
toric varieties}

\vspace{0.2in}

Ron Donagi$^1$, Josh Guffin$^1$, Sheldon Katz$^2$, Eric Sharpe$^3$ \\

$\,$

\begin{tabular}{cc}
{ \begin{tabular}{c}
$^1$ Department of Mathematics\\
University of Pennsylvania\\
David Rittenhouse Lab.\\
209 South 33rd St.\\
Philadelphia, PA  19104-6395
\end{tabular} } &
{ \begin{tabular}{c}
$^2$ Department of Mathematics \\
1409 W. Green St. \\
University of Illinois\\
Urbana, IL  61801
\end{tabular} } \\
{\begin{tabular}{c}
$^3$ Physics Department \\
Robeson Hall (0435) \\
Virginia Tech \\
Blacksburg, VA  24061
\end{tabular} }
\end{tabular} \\

$\,$

{\tt donagi@math.upenn.edu}, {\tt guffin@math.upenn.edu},
{\tt katz@math.uiuc.edu}, {\tt ersharpe@vt.edu} \\

$\,$

\end{center}

In this paper, we will outline computations of quantum sheaf cohomology
for deformations of tangent bundles of toric varieties, for those 
deformations describable as deformations of toric Euler sequences.
Quantum sheaf cohomology is a heterotic analogue of quantum cohomology,
a quantum deformation of the classical product on sheaf cohomology
groups, that computes nonperturbative
corrections to analogues of ${\bf \overline{27}}^3$
couplings in heterotic string compactifications.  Previous computations
have relied on either physics-based GLSM techniques or computation-intensive
brute-force Cech cohomology techniques.  This paper describes
methods for greatly simplifying mathematical computations, and derives
more general results than previously obtainable with GLSM techniques.
We will outline recent results (rigorous proofs will appear elsewhere).

\begin{flushleft}
October 2011
\end{flushleft}

\newpage

\tableofcontents

\newpage

\section{Introduction}

This paper is concerned with computing quantum sheaf cohomology
rings, an analogue of quantum cohomology rings for heterotic strings.

Quantum cohomology describes the operator product rings in
A model topological field theories.  Those operator product rings
are deformations of the classical cohomology rings, and so are called
`quantum cohomology' rings.  The deformations encode information about
minimal-area surfaces, and so quantum cohomology played an important
role in the enumerative geometry revolution that swept through algebraic
geometry starting in the early 1990s, and continues in various forms to
this day.

Quantum sheaf cohomology computes analogous invariants of
pairs consisting of spaces $X$ together with vector bundles
${\cal E} \rightarrow X$ satisfying the conditions
\begin{displaymath}
\Lambda^{\rm top} {\cal E}^* \: \cong \: K_X, \: \: \:
{\rm ch}_2({\cal E}) \: = \: {\rm ch}_2(TX).
\end{displaymath}
Such pairs define the `A/2 model,' a heterotic generalization of
the A model.  An analogue of quantum cohomology for the A/2 model
was originally defined in \cite{ks} (motivated by physics considerations
in \cite{abs}), and describes a deformation of the
product structure on sheaf cohomology, for which reason this
deformation has been named `quantum sheaf cohomology.'
Much as in ordinary quantum cohomology, the deformation in question
revolves around enumerative properties of $X$ -- specifically,
one computes sheaf cohomology of induced sheaves over a moduli space
of curves in $X$, corresponding physically to nonperturbative
corrections to correlation functions of charged fields.

Quantum sheaf cohomology and related notions
have been further developed in a variety of recent
papers including {\it e.g.}
\cite{gk,mcom,s2,s3,ade,tan1,tan2,ms,mm2,m1,kmmp,gs2,mp1,pmp,mcorev,g1,mm3}.

In this paper we shall outline general results for
quantum sheaf cohomology for $X$ a compact toric variety and
${\cal E}$ a deformation of the tangent bundle of $X$, described as a
deformation of the toric Euler sequence.
In particular, in the past such computations have been done with
either physics-based GLSM techniques (which so far have not been amenable
to studying nonlinear deformations), or math-based computation-intensive
brute-force Cech cohomology computations.  One of the innovations of this
paper and \cite{dgks} are a set of new ideas to radically simplify
mathematics computations, which we use to obtain results of greater
generality than previously obtainable with GLSM techniques.
Utilizing those methods, we find, for example,
that quantum sheaf cohomology rings, at least 
in these cases, are independent of nonlinear deformations, a result
previously conjectured in \cite{mm2,kmmp}.
Detailed proofs are left to \cite{dgks}.

We begin in section~\ref{sect:genlprocedure} by describing the
A/2 model (a holomorphic field theory), and outline the correlation
function computations in that theory, first at a formal level, then
describing generalities of linear sigma model 
(LSM) compactifications and induced sheaves
over moduli spaces of curves.  In section~\ref{sect:oneproj} we begin
by computing the quantum sheaf cohomology of a projective space.
Since the tangent bundle of a projective space is rigid, the result will
automatically match the ordinary quantum cohomology ring, but this is a
useful warm-up exercise and demonstration of some of the technology 
we are introducing that simplify general quantum sheaf cohomology
computations.
In section~\ref{sect:ex:proj}, 
we apply these ideas to compute quantum sheaf cohomology on a product
of projective spaces.  Briefly, quantum sheaf cohomology reduces to
a classical sheaf cohomology computation over the LSM moduli spaces,
and for a product of projective spaces, the LSM moduli spaces are again
a product of projective spaces, so we work through classical sheaf
cohomology for products of general projective spaces, then apply those
results to quickly compute quantum sheaf cohomology for a deformation
of the tangent bundle on ${\bf P}^1 \times {\bf P}^1$.
Projective spaces are a bit simple, so in section~\ref{sect:ex:hirz}
we compute quantum sheaf cohomology for a deformation of the tangent bundle
on a Hirzebruch surface, which allows us to tackle issues such as
nonlinear deformations and four-fermi interaction terms.  
In section~\ref{sect:genl} we describe general
results (derived in detail in \cite{dgks}).  In appendix~\ref{sect:glsm-4fermi}
we derive an ansatz for four-fermi terms from GLSM's, that is used both
in this paper and in \cite{dgks}.

\section{General procedure and definitions}
\label{sect:genlprocedure}

First, let us briefly review the A/2 model.
Recall that on the $(2,2)$ locus, the A model topological field
theory is a twist of the $(2,2)$ supersymmetric nonlinear sigma model 
\begin{displaymath}
\frac{1}{\alpha'} \int_{\Sigma} d^2z \left( 
         \left( g_{\mu \nu} +  i B_{\mu \nu} \right)
\partial \phi^{\mu} \overline{\partial} \phi^{\nu}
          \: + \:
\frac{i}{2} g_{\mu \nu} \psi_+^{\mu} D_{\overline{z}} \psi_+^{\nu} 
\: + \: \frac{i}{2} g_{\mu \nu} \psi_-^{\mu} D_z \psi_-^{\nu} 
\: + \: R_{i \overline{\jmath} k \overline{l}} 
\psi_+^i \psi_+^{\overline{\jmath}} \psi_-^k \psi_-^{\overline{l}}
\right),
\end{displaymath}
which is amenable
to rational curve counting.  Specifically, the A model is defined
by twisting worldsheet fermions into worldsheet scalars and vectors
as follows \cite{edtft}:
\begin{displaymath}
\begin{array}{lcl}
\psi_+^i \: \in \: \Gamma_{ C^{\infty} }\left( \phi^* T^{1,0}X \right), & \: \:
\: &
\psi_-^i \: \in \: \Gamma_{ C^{\infty} }\left(
\overline{K}_{\Sigma} \otimes \left( \phi^* T^{0,1} X \right)^{*} \right), \\
\psi_+^{\overline{\imath}} \: \in \:
\Gamma_{ C^{\infty} }\left( K_{\Sigma} \otimes \left(
\phi^* T^{1,0} X \right)^{*} \right), & \: \: \: &
\psi_-^{\overline{\imath}} \: \in \: \Gamma_{ C^{\infty} }\left(
\phi^* T^{0,1} X \right).
\end{array}
\end{displaymath}
The heterotic analogue of the A model, known as the A/2 model,
is a twist of the $(0,2)$ nonlinear
sigma model
\begin{displaymath}
 \frac{1}{\alpha'} \int_{\Sigma} d^2z \left(
\left(g_{\mu \nu} +  i B_{\mu \nu} \right) 
\partial \phi^{\mu} \overline{\partial} \phi^{\nu} 
\:+ \: \frac{i}{2} g_{\mu \nu} \psi_+^{\mu} D_{\overline{z}} \psi_+^{\nu}
\: + \: \frac{i}{2}h_{\alpha \beta} \lambda_-^{\alpha} D_{z} \lambda_-^{\beta}
\: + \: 
F_{i \overline{\jmath} a \overline{b}} \psi_+^i \psi_+^{\overline{\jmath}} 
\lambda_-^a \lambda_-^{\overline{b}}
\right),
\end{displaymath}
in which the fermions couple to bundles as follows:
\begin{displaymath}
\begin{array}{lcl}
\psi_+^i \: \in \: \Gamma_{ C^{\infty} }\left( \phi^* T^{1,0} X \right), &
\: \: \: & \lambda_-^a \: \in \: \Gamma_{ C^{\infty} }\left(
\overline{K}_{\Sigma} \otimes  \phi^* \overline{ \mathcal{E} }^*  
\right), \\
\psi_+^{\overline{\imath}} \: \in \: \Gamma_{ C^{\infty} } \left(
K_{\Sigma} \otimes \left( \phi^* T^{1,0}X \right)^{*} \right), 
& \: \: \: &
\lambda_-^{\overline{a}} \: \in \: \Gamma_{ C^{\infty} }\left(
\phi^* \overline{ \mathcal{E} } \right),
\end{array}
\end{displaymath}
where $\mathcal{E}$ is a holomorphic vector bundle on $X$.  Anomaly
cancellation requires
\begin{displaymath}
\Lambda^{\rm top} {\cal E}^* \: \cong \: K_X, \: \: \:
{\rm ch}_2({\cal E}) \: = \: {\rm ch}_2(TX).
\end{displaymath}
(The second statement is the Green-Schwarz anomaly cancellation condition
generic to all heterotic theories; the first is a condition specific to the
A/2 twist, an analogue of the condition that the closed string B model 
can only propagate on spaces $X$ such that $K_X^{\otimes 2}$ is trivial
\cite{s3,edtft}.)
In fact, a specific choice of isomorphism $\Lambda^{\rm top} {\cal E}^* \cong
K_X$ is part of the data needed to define the path integral.
Although both left- and right-movers have been twisted,
the theory defined by the twisting above is not a topological
field theory, since the worldsheet does not have supersymmetry
on left-movers.  Nevertheless it is sufficiently close to a 
true topological field theory to enable mathematical computations.

The RR states of the A/2 model generalizing the A model states
are counted by
sheaf cohomology $H^q(X, \Lambda^p \mathcal{E}^{\vee})$.

In general terms, we understand correlation functions in the A/2 model
as follows (see \cite{ks} for a more complete discussion).  
For a space $X$ with holomorphic vector bundle ${\cal E} 
\rightarrow X$ satisfying
\begin{displaymath}
\det {\cal E}^* \: \cong \: K_X, \: \: \:
{\rm ch}_2({\cal E}) \: = \: {\rm ch}_2(TX),
\end{displaymath}
the classical contribution to a correlation function is
\begin{displaymath}
\langle {\cal O}_1 \cdots {\cal O}_n \rangle \: = \:
\int_X \omega_1 \wedge \cdots \wedge \omega_n, 
\end{displaymath}
where each $\omega_i$ is an element of $H^*(X, \Lambda^* {\cal E}^*)$,
and corresponds to an operator ${\cal O}_i$.
The correlation function can only be nonzero if
\begin{displaymath}
\omega_1 \wedge \cdots \wedge \omega_n \: \in \: H^{\rm top}\left(X, 
\Lambda^{\rm top} {\cal E}^* \right)
\end{displaymath}
and we get a number from this because of the isomorphism
\begin{displaymath}
\det {\cal E}^* \: \cong \: K_X
\end{displaymath}
and the fact that $H^{\rm top}(X, K_X) \cong {\bf C}$.

In sectors of nonzero instanton degree, each ${\cal O}_i$ induces
an element of $H^*({\cal M}, \Lambda^* {\cal F}^*)$, where ${\cal M}$
is the moduli space and ${\cal F}$ a sheaf on ${\cal M}$ induced
by ${\cal E}$, as described in \cite{ks}.
For example,
if the moduli space ${\cal M}$ admitted a universal instanton $\alpha$,
then ${\cal F} = R^0 \pi_* \alpha^* {\cal E}$.
Schematically, if there are no $\psi_+^{\overline{\imath}}$,
$\lambda_-^a$ zero modes, then
the contribution to a correlation function in a sector of
nonzero instanton
degree will be of the form
\begin{displaymath}
\int_{\cal M} \tilde{\omega}_1 \wedge \cdots \wedge \tilde{\omega}_n,
\end{displaymath}
where each $\tilde{\omega}_i$ is an element of $H^*({\cal M},
\Lambda^* {\cal F}^*)$, and corresponds to an operator ${\cal O}_i$.
In close analogy with the classical case, this contribution will be
nonzero if
\begin{displaymath}
\tilde{\omega}_1 \wedge \cdots \wedge \tilde{\omega}_n
\: \in \: H^{\rm top}\left({\cal M}, \Lambda^{\rm top} {\cal F}^* \right)
\end{displaymath} 
and we get a number from this because the conditions
\begin{displaymath}
\det {\cal E}^* \: \cong \: K_X, \: \: \:
{\rm ch}_2({\cal E}) \: = \: {\rm ch}_2(TX)
\end{displaymath}
imply (via Grothendieck-Riemann-Roch) that
\begin{displaymath}
\det {\cal F}^* \: \cong \: K_{\cal M}.
\end{displaymath}

If there are $\psi_+^{\overline{\imath}}$, $\lambda_-^a$ zero modes,
then we have to make use of the four-fermi terms, as described in
\cite{ks}.  Define
\begin{displaymath}
{\cal F}_1 \: \equiv \: R^1 \pi_* \alpha^* {\cal E}, \: \: \:
{\rm Obs} \: \equiv \: R^1 \pi_* \alpha^* TX,
\end{displaymath}
then one can formally identify the contribution of each
four-fermi term with an insertion of
\begin{displaymath}
H^1\left( \mathcal{M}, \mathcal{F}^* \otimes \mathcal{F}_1 \otimes
\left( \mbox{Obs} \right)^* \right).
\end{displaymath}
Assuming equal numbers of $\psi_+^{\overline{\imath}}$, $\lambda_-^a$ zero
modes, correlation functions in such a sector will have the form
\begin{displaymath}
\int_{\cal M} \tilde{\omega}_1 \wedge \cdots \wedge \tilde{\omega}_n
\wedge \alpha,
\end{displaymath}
where the $\tilde{\omega}_i$ are as before, and $\alpha$ is a wedge product
of cohomology classes associated with four-fermi terms.
Altogether the contribution can only be nonzero if
\begin{displaymath}
 \tilde{\omega}_1 \wedge \cdots \wedge \tilde{\omega}_n
\wedge \alpha
\: \in \: H^{\rm top}\left( {\cal M}, \Lambda^{\rm top} {\cal F}^* \otimes
\Lambda^{\rm top} {\cal F}_1 \otimes \Lambda^{\rm top} {\rm Obs}^* \right)
\end{displaymath}
and we get a number from this because in these circumstances the conditions
\begin{displaymath}
\det {\cal E}^* \: \cong \: K_X, \: \: \:
{\rm ch}_2({\cal E}) \: = \: {\rm ch}_2(TX)
\end{displaymath}
imply (via Grothendieck-Riemann-Roch) that
\begin{displaymath}
\det {\cal F}^* \otimes \det {\cal F}_1 \otimes \det {\rm Obs}^*
\: \cong \: K_{\cal M}.
\end{displaymath}

Now, let us begin to specialize to examples of the form we shall
discuss in this paper.
Consider a projective toric variety $X = X_{\Sigma}$ over
${\bf C}$ of dimension $n$ with fan $\Sigma$.
The tangent bundle $TX$ is
defined by a cokernel of the form
\begin{displaymath}
0 \: \longrightarrow \: {\cal O}^{\oplus r} \: 
\stackrel{E}{\longrightarrow} \:
\bigoplus_{i=1}^n {\cal O}\left(\vec{q}_i\right)
\: \longrightarrow \: TX \: \longrightarrow \: 0,
\end{displaymath}
where $r$ is
the rank of the Picard lattice, whose complexification we
denote as 
\begin{displaymath}
W = {\rm Pic}(X) \otimes_{\bf Z} {\bf C}. 
\end{displaymath} 
We will often
denote ${\cal O}_X^{\oplus r}$ by $W \otimes {\cal O}_X$.
The map $E$ acts by mapping the $a^{\rm th}$ ${\cal O}$ as
\begin{displaymath}
{\cal O} \: \stackrel{ q_{ai} \phi_i }{\longrightarrow} \:
{\cal O}\left( \vec{q}_i \right),
\end{displaymath}
where $\phi_i \in \Gamma\left( {\cal O}\left( \vec{q}_i \right) \right)$
is a homogeneous coordinate on the toric variety (see for
example \cite{cls}).

Now, we shall consider deformations ${\cal E}$ of the tangent bundle
above, defined by cokernels
\begin{displaymath}
0 \: \longrightarrow \: {\cal O}^{\oplus r} \: \stackrel{E}{\longrightarrow} \:
\bigoplus_{i=1}^n {\cal O}\left(\vec{q}_i\right)
\: \longrightarrow \: {\cal E} \: \longrightarrow \: 0
\end{displaymath}
for more general maps $E$.
Each element of $E$ will be a polynomial.  We will distinguish two types
of contributions to $E$:  ``linear'' and ``nonlinear'' deformations.
Linear deformations involve monomials containing a single homogeneous
coordinate (as in all of the maps defining the tangent bundle).
Nonlinear deformations involve monomials containing a product of more than
one homogeneous coordinate.

We will use the `linear sigma model' moduli space ${\cal M}$.
As explained in {\it e.g.} \cite{dr}, for the case above,
this is constructed by expanding each of the homogeneous coordinates
on $X$ in a basis of zero modes on ${\bf P}^1$, and interpreting the
coefficients in the expansion as homogeneous coordinates on the moduli space.
If 
\begin{displaymath}
X \: = \: {\bf C}^n // \left( {\bf C}^{\times} \right)^r,
\end{displaymath}
where $({\bf C}^{\times})^r$ acts on ${\bf C}^n$ with weights
$\vec{q}_i$, then the linear sigma model moduli space of maps of 
degree $\vec{d}$ is
\begin{displaymath}
{\cal M} \: = \: \left( \oplus_{i=1}^n H^0\left( {\bf P}^1,
{\cal O}(\vec{q}_i \cdot \vec{d}) \right) \right) //
\left( {\bf C}^{\times} \right)^r.
\end{displaymath}

It can be shown that the LSM moduli space ${\cal M}$ is smooth
whenever the original toric variety is.  (The basic point is that if we
describe the toric variety as $({\bf C}^n - E)/G$, then singularities are at
fixed points of $G$.  See \cite{dgks}[section 4.1] for further details.)

The induced sheaves ${\cal F}$, ${\cal F}_1$ can be constructed
in an analogous fashion \cite{ks}, by expanding worldsheet GLSM fermions
in a basis of zero modes and interpreting the coefficients as line bundles
over the moduli space.  Specifically, following the methods
of \cite{ks}, one finds for present case that
\begin{displaymath}
0 \: \longrightarrow \: {\cal O}^{\oplus r} \: 
\stackrel{E'^T}{\longrightarrow} \:
\bigoplus_{i=1}^n H^0\left( {\bf P}^1, {\cal O}(\vec{q}_i \cdot
\vec{d}) \right) \otimes_{ {\bf C} } {\cal O}(\vec{q}_i )
\: \longrightarrow \: {\cal F} \: \longrightarrow \: 0,
\end{displaymath}
\begin{displaymath}
{\cal F}_1 \: \cong \: \bigoplus_{i=1}^n H^1\left( {\bf P}^1,
{\cal O}(\vec{q}_i \cdot \vec{d} ) \right) \otimes_{ {\bf C} }
{\cal O}(\vec{q}_i).
\end{displaymath}

The map $E'$ in the definition of ${\cal F}$ is induced from the corresponding
map in the definition of ${\cal E}$.  It is constructed by taking the
map $E$ in ${\cal E}$ (a polynomial in homogeneous coordinates) and expanding
in terms of homogeneous coordinates on the worldsheet ${\bf P}^1$.
The components of the induced map $E'$ are then the coefficients of
various monomials in the homogeneous coordinates on ${\bf P}^1$.

\begin{figure}[h]
      \begin{center}
         \begin{picture}(150,150)
            \Line(75,75)(150,75)
            \Line(75,0)(75,150)
            \Line(0,110)(75,75)
            \Text(145,80)[b]{$u$}
            \Text(80,145)[l]{$s$}
            \Text(80,5)[l]{$t$}
            \Text(5,110)[b]{$v$}
            \Text(5,100)[t]{$(-1,n)$}
         \end{picture}
      \end{center}
      \caption{The fan for ${\bf F}_n$}
      \label{fig:FnFan}
   \end{figure}
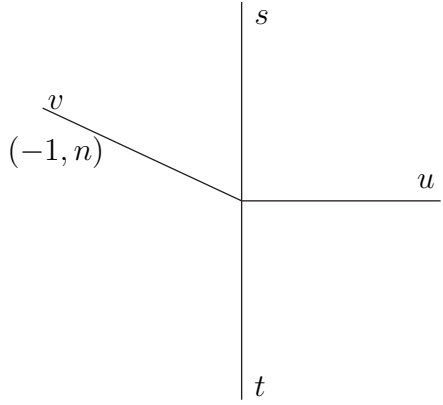

To explain how the map is induced in more detail, let us consider
the example of a Hirzebruch surface ${\bf F}_n$.
 To set notation, describe the Hirzebruch surface by the toric fan
   in Figure \ref{fig:FnFan}.

From the fan, we read off the relations between toric divisors
\begin{displaymath}
D_u \: = \: D_v, \: \: \:
D_t \: = \: D_s \: + \: n D_v
\end{displaymath}
and the Stanley-Reisner ideal
\begin{displaymath}
D_u \cdot D_v \: = \: 0 \: = \:
D_s \cdot D_t.
\end{displaymath}
The homogeneous coordinates $u$, $v$, $s$, $t$ (corresponding to the
four toric divisors) have the following weights under two ${\bf C}^{\times}$
actions:
\begin{center}
\begin{tabular}{cccc}
$u$ & $v$ & $s$ & $t$ \\ \hline
$1$ & $1$ & $0$ & $n$ \\
$0$ & $0$ & $1$ & $1$
\end{tabular}
\end{center}
We describe a deformation $\cE^*$ of the cotangent bundle as the
cokernel
\begin{displaymath}
0 \: \longrightarrow \: \cE^* \: \longrightarrow \:
{\mathcal O}(-1,0)^{\oplus 2} \oplus {\mathcal O}(0,-1) \oplus
{\mathcal O}(-n,-1) \: \stackrel{E}{\longrightarrow} \: W \otimes
{\mathcal O} \: \longrightarrow \: 0,
\end{displaymath}
where $W$ is a two-dimensional vector space,
\begin{equation}  \label{hirz-genl-map}
E \: = \: \left[ \begin{array}{cc}
A x & B x \\
\gamma_1 s & \gamma_2 s \\
\alpha_1 t + s f_1(u,v) & \alpha_2 t + s f_2(u,v)
\end{array} \right],
\end{equation}
with
\begin{displaymath}
x \: \equiv \: \left[ \begin{array}{c} u \\ v \end{array} \right],
\end{displaymath}
$A$, $B$ constant $2 \times 2$ matrices, $\gamma_1$, $\gamma_2$,
$\alpha_1$, $\alpha_2$ constants, and $f_{1,2}(u,v)$ homogeneous polynomials
of degree $n$.  (The matrices $A$, $B$, and $\gamma_{1,2}$ define
linear deformations of the tangent bundle; the functions
$sf_{1,2}(u,v)$ define nonlinear deformations.)

To demonstrate the technology, consider for a moment maps of degree
$(1,0)$.  In this case, we get the induced sheaf
\begin{displaymath}
0 \: \longrightarrow \: {\cal F}^* \: \longrightarrow \:
{\cal O}(-1,0)^4 \oplus {\cal O}(0,-1) \oplus {\cal O}(-n,-1)^{n+1}
\: \stackrel{E'}{\longrightarrow} \: W \otimes {\cal O} \: \longrightarrow \: 0,
\end{displaymath}
where the map $E'$ is induced from the map $E$ by
expanding fields in zero modes and picking off terms with the same
homogeneous coordinates on ${\bf P}^1$.  Let us work through that in
detail to illustrate the result.
In the degree $(1,0)$ sector, we expand
\begin{eqnarray*}
u & = & u_0 a \: + \: u_1 b, \\
v & = & v_0 a \: + \: v_1 b, \\
s & = & s_0, \\
t & = & t_0 a^n \: + \: t_1 a^{n-1} b \: + \: \cdots \: + \: t_n b^n,
\end{eqnarray*}
where $a$, $b$ are homogeneous coordinates on ${\bf P}^1$.
Then, in the original map $E$, we replace each field $u$, $v$, $s$, $t$
by its expansion in zero modes above, and pick off terms with the
same homogeneous coordinates.
In this fashion, we find
\begin{equation}  \label{fullzeromodeexp}
E' \: = \: \left[
\begin{array}{cc}
\left[ \begin{array}{cc}
A & 0 \\ 0 & A \end{array} \right] x' &
\left[ \begin{array}{cc}
B & 0 \\ 0 & B \end{array} \right] x' \\[1em]
\gamma_1 s_0 & \gamma_2 s_0 \\[1em]
\alpha_1 t_0 \: + \: s_0 f_{10} u_0^n \: + \: s_0 f_{11} u_0^{n-1} v_0
&
\alpha_2 t_0 \: + \: s_0 f_{20} u_0^n \: + \: s_0 f_{21} u_0^{n-1} v_0
 \\
\: + \: \cdots \: + \: s_0 f_{1n} v_0^n
&
\: + \: \cdots \: + \: s_0 f_{2n} v_0^n \\[1em]
\alpha_1 t_1 \: + \: s_0 f_{10} (n u_0^{n-1} u_1) 
\: + \: s_0 f_{11} u_0^{n-1} v_1
& \alpha_2 t_1 \: + \: s_0 f_{20} (n u_0^{n-1} u_1) 
\: + \: s_0 f_{21} u_0^{n-1} v_1
 \\
 \: + \: (n-1) s_0 f_{11} u_0^{n-2} u_1
v_0  \: + \: \cdots
&  \: + \: (n-1) s_0 f_{21} u_0^{n-2} u_1
v_0  \: + \: \cdots
 \\[1em]
\cdots & \cdots
\end{array}
\right],
\end{equation}
where
\begin{displaymath}
x' \: = \: [ u_0, v_0, u_1, v_1 ]^T 
\end{displaymath}
and
\begin{displaymath}
f_i(u,v) \: = \: f_{i0} u^n \: + \: f_{i1} u^{n-1} v \: + \: \cdots
\: + \: f_{in} v^n.
\end{displaymath}
In $E'$, the lines with $t_0$, for example, correspond to coefficients of
$a^n$, the lines with $t^1$ correspond to coefficients of $a^{n-1} b$,
and so forth.

It can be shown in general that ${\cal F}$ is locally-free whenever
${\cal E}$ is locally-free \cite{dgks}.  Briefly, ${\cal F}$ will be
locally-free whenever $E'$ is surjective.  At any point on the GLSM moduli
space, pick a point on ${\bf P}^1$ at which the corresponding map
is nondegenerate, then the image of $E'$ is the image of $E$,
hence surjectivity of $E$ implies surjectivity of $E'$.

\section{Example:  projective space}
\label{sect:oneproj}

Let us begin with an extremely simple example, namely
${\bf P}^n$.  We will consider what appears formally to be a deformation
of the tangent bundle of ${\bf P}^n$, defined by ${\cal E}$ below:
\begin{displaymath}
0 \: \longrightarrow \: {\cal E}^* \: \longrightarrow \: Z_0 \:
\stackrel{E}{\longrightarrow} \: W \otimes {\cal O} \: \longrightarrow \: 0,
\end{displaymath}
where
\begin{displaymath}
Z_0 \: = \: {\cal O}(-1)^{\oplus n+1}, \: \: \: E \: = \: A x,
\end{displaymath}
where $W$ is a one-dimensional vector space,
$x$ is a vector of homogeneous coordinates on ${\bf P}^n$, and
$A$ a constant $(n+1) \times (n+1)$ matrix.
We say this appears to be a deformation; however, the tangent bundle of
${\bf P}^n$ admits no deformations, hence the matrix $A$ encodes,
for nondegenerate $A$, mere reparametrizations.  By contrast,
for ${\bf P}^1\times {\bf P}^1$, which we shall study in the next
section, generic deformations of the tangent bundle yield bundles
which are not isomorphic to the original tangent bundle.

Since we are simply giving a more complicated description of ${\bf P}^n$
with its tangent bundle, the quantum sheaf cohomology ring we compute should
exactly match the ordinary quantum cohomology ring, which is what we shall
find.  This example will serve as a useful computational exercise,
but we will not start generating new results until the next section.

First, let us consider the classical cohomology ring.
A nonzero correlation function arises from  
correlators of total degree $n$, equal to the dimension of ${\bf P}^n$.
Classical correlation functions are then a map
\begin{displaymath}
{\rm Sym}^n W \: = \: H^0\left( {\rm Sym}^n W \otimes {\cal O} \right)
\: \longrightarrow \: H^n\left(\Lambda^n {\cal E}^* \right).
\end{displaymath}
To determine the map, we use the generalized Koszul complex associated
to $\Lambda^n 
{\cal E}^*$:
\begin{displaymath}
0 \: \longrightarrow \: \Lambda^n {\cal E}^* \: \longrightarrow \:
\Lambda^n Z_0 \: \longrightarrow \: \Lambda^{n-1} Z_0 \otimes W \:
\longrightarrow \: \cdots \: \longrightarrow \: {\rm Sym}^n W \otimes
{\cal O} \: \longrightarrow \: 0,
\end{displaymath}
which factorizes into a series of maps
\begin{equation} \label{pn:1}
0 \: \longrightarrow \: \Lambda^n {\cal E}^* \: \longrightarrow \:
\Lambda^n Z_0 \: \longrightarrow \: S_{n-1} \: \longrightarrow \: 0,
\end{equation}
\begin{equation}  \label{pn:2}
0 \: \longrightarrow \: S_i \: \longrightarrow \: \Lambda^i Z_0 \otimes
{\rm Sym}^{n-i} W \: \longrightarrow \: S_{i-1} \: \longrightarrow \: 0,
\end{equation}
\begin{equation}  \label{pn:3}
0 \: \longrightarrow \: S_1 \: \longrightarrow \: Z \otimes
{\rm Sym}^{n-1} W \: \longrightarrow \: {\rm Sym}^n W \otimes {\cal O}
\: \longrightarrow \: 0.
\end{equation}
Now, $H^j(\Lambda^i Z_0)$ will vanish unless $j=n$, $i=n+1$
(or $i=j=0$, but we shall suppress that case as it will not be pertinent
for our computations).   
Thus, from~(\ref{pn:1}) we find
\begin{displaymath}
H^n\left( \Lambda^n {\cal E}^* \right) \: \stackrel{\sim}{\longrightarrow}
\: H^{n-1}(S_{n-1}), 
\end{displaymath}
from~(\ref{pn:2}) we find
\begin{displaymath}
H^{i-1}(S_{i-1}) \: \stackrel{\sim}{\longrightarrow} \: H^i(S_i)
\end{displaymath}
for $2 \leq i \leq n-1$, 
and from~(\ref{pn:3}) we find
\begin{displaymath}
H^0({\rm Sym}^n W \otimes {\cal O}) \: \stackrel{\sim}{\longrightarrow} \:
H^1(S_1),
\end{displaymath}
which implies that the map
\begin{displaymath}
{\rm Sym}^n W \: \stackrel{\sim}{\longrightarrow} \: H^n(\Lambda^n {\cal E}^*)
\end{displaymath}
is an isomorphism, as indicated.  
(As a consistency check, note that since $W$ is
one-dimensional, ${\rm Sym}^n W$ is also one-dimensional.)

Now, in principle the factorization of the generalized
Koszul complex will
stop giving isomorphisms if ever we need to compute $H^n(\Lambda^{n+1} Z_0)$.
This will happen if we consider correlation functions with correlators of
degree greater than $n$.  For example, if we have degree $n+1$ correlators,
then the correlation function computes a map
\begin{displaymath}
{\rm Sym}^{n+1} W \: \longrightarrow \: H^{n+1}(\Lambda^{n+1} {\cal E}^*)
\: = \: 0.
\end{displaymath}
In this (trivial) case, we have the generalized Koszul complex
\begin{displaymath}
0 \: \longrightarrow \: \Lambda^{n+1} {\cal E}^* (=0) \:
\longrightarrow \: \Lambda^{n+1} Z_0 \: \longrightarrow \: 
\Lambda^n Z_0 \otimes W \: \longrightarrow \: \cdots \: \longrightarrow \:
{\rm Sym}^{n+1} W \otimes {\cal O} \: \longrightarrow \: 0,
\end{displaymath}
which factorizes as
\begin{equation}  \label{pna:1}
0 \: \longrightarrow \: \Lambda^{n+1} {\cal E}^* (=0) \:
\longrightarrow \: \Lambda^{n+1} Z_0 \: \longrightarrow \: S_n \:
\longrightarrow \: 0,
\end{equation}
\begin{equation}  \label{pna:2}
0 \: \longrightarrow \: S_i \: \longrightarrow \: \Lambda^i Z_0 \otimes
{\rm Sym}^{n+1-i} W \: \longrightarrow \: S_{i-1} \: \longrightarrow \: 0,
\end{equation}
\begin{equation} \label{pna:3}
0 \: \longrightarrow \: S_1 \: \longrightarrow \: Z \otimes {\rm Sym}^n W
\: \longrightarrow \: {\rm Sym}^{n+1} W \otimes {\cal O} \: \longrightarrow
\: 0.
\end{equation}
As before, 
from~(\ref{pna:3}) we have
\begin{displaymath}
H^0\left( {\rm Sym}^{n+1} W \otimes {\cal O} \right) \: 
\stackrel{\sim}{\longrightarrow} \: H^1(S_1)
\end{displaymath}
and from~(\ref{pna:2}) we have
\begin{displaymath}
H^{i-1}(S_{i-1}) \: \stackrel{\sim}{\longrightarrow} \: H^i(S_i)
\end{displaymath}
for $2 \leq i \leq n$.
Finally from~(\ref{pna:1}) we have
\begin{displaymath}
H^n(S_n) \: \stackrel{\sim}{\longrightarrow} \: H^n(\Lambda^{n+1} Z_0).
\end{displaymath}
Thus, the original correlation function necessarily vanishes:
\begin{displaymath}
{\rm Sym}^{n+1} W \: \longrightarrow \: H^{n+1}(\Lambda^{n+1} {\cal E}^*) = 0
\end{displaymath}
but 
\begin{displaymath}
{\rm Sym}^{n+1} W \: \stackrel{\sim}{\longrightarrow} \:
H^n(\Lambda^{n+1} Z_0).
\end{displaymath}
From \cite{dgks}[section 3.3], the group $H^n(\Lambda^{n+1} Z_0)$
is a one-dimensional vector space generated by
\begin{displaymath}
\det (A \psi) \: = \: \left( \det A \right) \psi^{n+1},
\end{displaymath}
where $\psi$ is a basis element for $W$.

Thus, we find the classical sheaf cohomology ring is of the form
\begin{displaymath}
{\bf C}[\psi] / \left( \det(A \psi) \right) \: \cong \:
{\bf C}[\psi] / \left( \psi^{n+1} \right).
\end{displaymath}

Now, let us turn to the quantum sheaf cohomology ring.
We shall compute this by first computing the (classical) sheaf
cohomology ring in any sector of fixed instanton degree, then relating
sectors of different instanton number.

In a sector of instanton number $d$, the linear sigma model moduli space
of ${\bf P}^n$ is easily computed to be ${\bf P}^{(n+1)(d+1) -1}$.
The induced bundle over the LSM moduli space is ${\cal F}$, where
\begin{displaymath}
0 \: \longrightarrow \: {\cal F}^* \: \longrightarrow \:
Z \: \stackrel{E'}{\longrightarrow} \: W \otimes {\cal O} \:
\longrightarrow \: 0,
\end{displaymath}
where
\begin{displaymath}
Z \: = \: {\cal O}(-1)^{\oplus (n+1)(d+1)},
\end{displaymath}
$W$ is the same one-dimensional vector space from previously, and
\begin{displaymath}
E' \: = \: \left[ \begin{array}{c}
A x_0 \\
A x_1 \\
\vdots \\
A x_d  \end{array} \right],
\end{displaymath}
where $x_i$ is a $(n+1)$-element vector of coefficients of fixed degree
in the expansion of homogeneous coordinates of ${\bf P}^n$ in zero modes.

We can now re-use the classical results above.  For fixed instanton degree $d$,
the sheaf cohomology ring is
\begin{displaymath}
{\bf C}[\psi] / \left( ( \det (A \psi) )^{d+1} \right) \: \cong \:
{\bf C}[\psi] / \left( \psi^{(n+1)(d+1)} \right).
\end{displaymath}
Therefore, to preserve kernels, any relation between correlation functions
in different sectors of fixed degree must be generated by
\begin{displaymath}
\langle {\cal O} \rangle_{0} \: \propto \:
\langle {\cal O} \left( \det( A \psi) \right)^{d} \rangle_d.
\end{displaymath}
(This ensures that if ${\cal O}$ is an element of the quotiented ideal
in the zero-degree sector, so that $\langle {\cal O} \rangle_0$ vanishes,
then its image ${\cal O} ( \det (A \psi) )^d$ will be an element of the
quotiented ideal in the sector of degree $d$, so that the corresponding
correlation function also vanishes.)

The relation above then implies that
\begin{displaymath}
\det (A \psi) \: = \: q
\end{displaymath}
for some constant $q$.  (For example, this follows immediately in
$d=1$, then higher degrees must just be a power.)
Since $\det(A \psi) = ( \det A ) \psi$, this is equivalent to the
relation
\begin{displaymath}
\psi^{n+1} \: = \: q',
\end{displaymath}
which is the standard quantum cohomology relation for a projective
space ${\bf P}^n$.

In the next sections, our tangent bundle deformations will, in general,
yield bundles that are not isomorphic to the tangent bundle, so the
quantum sheaf cohomology relations will be nontrivial.

\section{Example:  product of projective spaces}
\label{sect:ex:proj}

Mathematical computations of quantum sheaf cohomology have
previously \cite{ks,gk,s2,g1} relied on brute-force Cech cohomology
representations.  One of the advancements of this paper and \cite{dgks}
is the use of purely analytic methods to derive quantum sheaf cohomology.

We will illustrate these advances through an explicit computation
for general deformations of the tangent bundle of ${\bf P}^1 \times
{\bf P}^1$.  In particular, previously special deformations of
the tangent bundle of ${\bf P}^1 \times {\bf P}^1$ have been computed
with brute-force Cech techniques, so this seems an appropriate
example to generalize here.  We will begin by examining classical cup
products for ${\bf P}^1 \times {\bf P}^1$, then classical
cup products for ${\bf P}^n \times {\bf P}^m$, and then we will
describe the quantum sheaf cohomology ring for ${\bf P}^1 \times
{\bf P}^1$, which will ultimately be determined by the classical
computations on products of more general projective spaces.

\subsection{Classical cup products on ${\bf P}^1 \times {\bf P}^1$}
\label{classical-p12-sect}

In this section we will discuss how to compute classical cup
products in the sheaf cohomology, without having to work through
a Cech cohomology computation.

Define $V = \Gamma( {\cal O}(1,0) )$, $\tilde{V} = \Gamma( {\cal O}(0,1) )$,
$W = {\bf C}^2$.
Define
\begin{displaymath}
Z_0 \: \equiv \:
\left( V \otimes {\cal O}(-1,0) \right) \oplus
\left( \tilde{V} \otimes {\cal O}(0,-1) \right).
\end{displaymath}
Then, the cotangent bundle deformation ${\cal E}^*$ is the kernel
\begin{displaymath}
0 \: \longrightarrow \: {\cal E}^* \: \longrightarrow \:
Z_0 \: \stackrel{E}{\longrightarrow} \: W \otimes {\cal O}
\: \longrightarrow \: 0,
\end{displaymath}
where
\begin{displaymath}
E \: = \: \left[
\begin{array}{cc}
Ax & Bx \\
C\tilde{x} & D\tilde{x}
\end{array} \right],
\end{displaymath}
where $x$, $\tilde{x}$ are vectors of homogeneous coordinates on
each ${\bf P}^1$ factor.

First, let us compute the classical sheaf cohomology ring.
Classical correlation functions are a map\footnote{
More formally, we could think of classical correlation functions and the
map above as an element of
\begin{displaymath}
{\rm Ext}^2\left( {\rm Sym}^2 \left( W \otimes {\cal O} \right),
\Lambda^2 {\cal E}^* \right),
\end{displaymath}
which corresponds to the exact sequence~(\ref{longex-l2e*}).
Breaking that long sequence into two short exact sequences along $Q$
corresponds to writing the Ext element above as a product of elements
of
\begin{displaymath}
{\rm Ext}^1\left( Q, \Lambda^2 {\cal E}^* \right),
\: \: \:
{\rm Ext}^1\left( {\rm Sym}^2 \left( W \otimes {\cal O} \right), Q \right),
\end{displaymath}
which correspond to the short exact sequences~(\ref{first-break}),
(\ref{second-break}).
}
\begin{displaymath}
{\rm Sym}^2 W \: = \: H^0\left( {\rm Sym}^2 W \otimes {\cal O} \right)
\: \longrightarrow \: H^2\left( \Lambda^2 {\cal E}^* \right)
\end{displaymath}
and we will determine the ring structure by computing the kernel of that
map.
We will use the generalized Koszul complex of $\Lambda^2 {\cal E}^*$:
\begin{equation}  \label{longex-l2e*}
0 \: \longrightarrow \: \Lambda^2 {\cal E}^* \:
\longrightarrow \:
\Lambda^2 Z_0 \: \longrightarrow \: Z_0 \otimes \left(
W \otimes {\cal O} \right) \: \longrightarrow \:
{\rm Sym}^2 \left( W \otimes {\cal O} \right) \: \longrightarrow \:
0.
\end{equation}

It remains to compute the cup product above.
First, split the long exact sequence~(\ref{longex-l2e*}) into
a pair of short exact sequences:
\begin{equation}  \label{first-break}
0 \: \longrightarrow \: \Lambda^2 {\cal E}^* \: \longrightarrow \:
\Lambda^2 Z_0 \: \longrightarrow \: Q \: \longrightarrow \: 0, 
\end{equation}
\begin{equation}   \label{second-break}
0 \: \longrightarrow \: Q \: \longrightarrow \:
Z_0 \otimes \left( W \otimes {\cal O} \right) \: \longrightarrow \:
{\rm Sym}^2\left( W \otimes {\cal O} \right) \:
\longrightarrow \: 0,
\end{equation}
which define $Q$.

Next, we shall evaluate the kernel of that map, that product,
which will give us the classical sheaf cohomology ring structure.

The short exact sequence~(\ref{second-break}) induces a map
\begin{displaymath}
\delta_1: \:
H^0\left( {\rm Sym}^2 W \otimes {\cal O} \right) \: \longrightarrow
\: H^1\left( Q \right)
\end{displaymath}
(from the associated long exact sequence).  Moreover, because
\begin{displaymath}
H^*\left(Z_0 \otimes W \right) \: = \: 0
\end{displaymath}
the map $\delta_1$ above is an isomorphism.

The other short exact sequence, (\ref{first-break}), induces
\begin{displaymath}
0 \: \longrightarrow \:
H^1\left( \Lambda^2 Z_0 \right) \: \longrightarrow \:
H^1(Q) \: \stackrel{\delta_2}{\longrightarrow} \:
H^2\left( \Lambda^2 {\cal E}^* \right)
\: \longrightarrow \: 0,
\end{displaymath}
using the fact that
\begin{displaymath}
H^1\left(\Lambda^2 {\cal E}^* = K_{ {\bf P}^1 \times {\bf P}^1 } \right)
\: = \: 0, \: \: \:
H^2\left( \Lambda^2 Z_0 \right) \: = \: 0.
\end{displaymath}

The classical cup product is then the composition
\begin{equation}   \label{classical-cup-product-map}
H^0\left( {\rm Sym}^2 W \otimes {\cal O} \right) 
\: \stackrel{\delta_1}{\longrightarrow} \:
H^1(Q) \: \stackrel{\delta_2}{\longrightarrow} 
\: H^2\left( \Lambda^2 {\cal E}^* \right).
\end{equation}
We have seen that $\delta_1$ is an isomorphism, but $\delta_2$ has
a nontrivial kernel.  Specifically, since
\begin{displaymath}
\Lambda^2 Z_0 \: = \: \left( \Lambda^2 V \otimes {\cal O}(-2,0) \right)
\oplus \left( \Lambda^2 \tilde{V} \otimes {\cal O}(0,-2) \right) \oplus
\left( V \otimes \tilde{V} \otimes {\cal O}(-1,-1) \right),
\end{displaymath}
we see that the kernel of the classical cup product is two-dimensional:
\begin{displaymath}
 H^1\left( \Lambda^2 Z_0 \right) \: = \:
\Lambda^2 V \oplus \Lambda^2 \tilde{V}.
\end{displaymath}
In fact, it can be shown \cite{dgks}[section 3.3] that the kernel
of the cup product map~(\ref{classical-cup-product-map}) 
is defined by the relations
\begin{eqnarray}
\det\left( \psi A \: + \: \tilde{\psi}B \right) & = & 0, 
\label{classical-p12-a}\\
\det\left( \psi C \: + \: \tilde{\psi}D \right) & = & 0.
\label{classical-p12-b}
\end{eqnarray}
These are the classical sheaf cohomology ring relations.

Let us check that this correctly reproduces the results of
\cite{gk}.  In that paper, $A = D = I$,
\begin{displaymath}
B \: = \: \left[ \begin{array}{cc}
\epsilon_1 & \epsilon_2 \\
\epsilon_3 & 0 
\end{array} \right], \: \: \:
C \: = \: \left[ \begin{array}{cc}
\gamma_1 & \gamma_2 \\
\gamma_3 & 0
\end{array} \right].
\end{displaymath}
There, the classical cohomology ring is given by
\begin{eqnarray*}
\psi^2 \: + \: \epsilon_1 \psi \tilde{\psi} \: - \:
\epsilon_2 \epsilon_3 \tilde{\psi}^2 & = & 0, \\
\tilde{\psi}^2 \: + \: \gamma_1 \psi \tilde{\psi} \: - \:
\gamma_2 \gamma_3 \psi^2 & = & 0.
\end{eqnarray*}
Applying the general methods above to the matrices $A$, $B$, $C$, $D$ here,
we find that
\begin{eqnarray*}
\det\left( \psi A \: + \: \tilde{\psi}B \right) & = & 
\psi^2 \: + \: \epsilon_1 \psi \tilde{\psi} \: - \:
\epsilon_2 \epsilon_3 \tilde{\psi}^2, \\
\det\left( \psi C \: + \: \tilde{\psi}D \right) & = &
\tilde{\psi}^2 \: + \: \gamma_1 \psi \tilde{\psi} \: - \:
\gamma_2 \gamma_3 \psi^2,
\end{eqnarray*}
and so we recover the results of \cite{gk} for the classical cohomology
ring as a special case.  Similarly, it is straightforward to check
that this also agrees with the general results of \cite{mcom}, as we
shall review later in section~\ref{sect:genl}.

\subsection{Classical cup products on ${\bf P}^n \times {\bf P}^m$}

Let us now quickly repeat the analysis of the previous subsection for
a more general product of projective spaces, ${\bf P}^n \times {\bf P}^m$.
In the next section, we will compute the quantum sheaf cohomology ring
for ${\bf P}^1 \times {\bf P}^1$, which will be determined by
classical computations on ${\bf P}^n \times {\bf P}^m$.

As before, define $V = \Gamma({\cal O}(1,0))$, 
$\tilde{V} = \Gamma({\cal O}(0,1))$, $W = {\bf C}^2$.  Define
\begin{displaymath}
Z \: = \: \left( V \otimes {\cal O}(-1,0) \right) \oplus \left(
\tilde{V} \otimes {\cal O}(0,-1) \right).
\end{displaymath}
Then, as before, the cotangent bundle deformation ${\cal E}^*$ is the
kernel
\begin{displaymath}
0 \: \longrightarrow \: {\cal E}^* \: \longrightarrow \: Z \: 
\stackrel{E}{\longrightarrow} \: W \otimes {\cal O} \: \longrightarrow \: 0,
\end{displaymath}
where
\begin{displaymath}
E \: = \: \left[ \begin{array}{cc}
\tilde{A} x & \tilde{B} x \\
\tilde{C} \tilde{x} & \tilde{D} \tilde{x}
\end{array} \right],
\end{displaymath}
where $x$, $\tilde{x}$ are vectors of homogeneous coordinates on
${\bf P}^n$, ${\bf P}^m$, respectively, $\tilde{A}$, $\tilde{B}$ are
$(n+1)\times(n+1)$ matrices, and $\tilde{C}$, $\tilde{D}$ are
$(m+1)\times(m+1)$ matrices.

As before, we think of classical correlation functions in this theory
as maps
\begin{displaymath}
{\rm Sym}^{n+m} W \: \longrightarrow \: H^{n+m}\left( \Lambda^{\rm top} 
{\cal E}^* \right)
\end{displaymath}
and we compute the kernel, using the generalized Koszul complex
associated to
$\Lambda^{\rm top} {\cal E}^*$:
\begin{displaymath}
0 \: \longrightarrow \: \Lambda^{n+m} {\cal E}^* \: 
\longrightarrow \: \Lambda^{n+m} Z \: \longrightarrow \:
\Lambda^{n+m-1} Z \otimes W \: \longrightarrow \cdots \longrightarrow
\: {\rm Sym}^{n+m} W \otimes {\cal O} \: \longrightarrow \: 0.
\end{displaymath}
To do computations, we split this into short exact sequences:
\begin{equation}  \label{pnpma}
0 \: \longrightarrow \: \Lambda^{n+m}{\cal E}^* \: \longrightarrow \:
\Lambda^{n+m}Z \: \longrightarrow \: S_{n+m-1} \: \longrightarrow \: 0,
\end{equation}
\begin{equation}   \label{pnpmb}
0 \: \longrightarrow \: S_i \: \longrightarrow \:
\Lambda^i Z \otimes {\rm Sym}^{n+m-i} W \: \longrightarrow \: S_{i-1}
\: \longrightarrow \: 0,
\end{equation}
\begin{equation}    \label{pnpmc}
0 \: \longrightarrow \: S_1 \: \longrightarrow \:
Z \otimes {\rm Sym}^{n+m-1} W \: \longrightarrow \:
{\rm Sym}^{n+m} W \otimes {\cal O} \: \longrightarrow \: 0.
\end{equation}

Now, $H^j(\Lambda^i Z)$ will vanish unless $j=i-1=n,m$ (see for
example \cite{dgks}) (or, alternatively, if $i=j=0$, but we shall
suppress that case as it will not be pertinent for our computations).  
Thus, from~(\ref{pnpma}), we find
\begin{displaymath}
H^{n+m-1}(S_{n+m-1}) \: \stackrel{\sim}{\longrightarrow} \:
H^{n+m}(\Lambda^{n+m}{\cal E}^*),
\end{displaymath}
from~(\ref{pnpmc}) we find
\begin{displaymath}
H^0({\rm Sym}^{n+m} W \otimes {\cal O}) \: \stackrel{\sim}{\longrightarrow}
\: H^1(S_1),
\end{displaymath}
and from~(\ref{pnpmb}) we find a surjective map
\begin{displaymath}
H^{i-1}(S_{i-1}) \: \longrightarrow \: H^i(S_i)
\end{displaymath}
for $2 \leq i \leq n+m-1$.
If $i-1 \neq n,m$, then the surjective map above is an isomorphism.
If $i-1$ is either $n$ or $m$, then it has a nontrivial kernel,
given by $H^{i-1}(\Lambda^i Z \otimes {\rm Sym}^{n+m-i}W)$.

If $i-1=n\neq m$, then $H^{i-1}(\Lambda^i Z) \cong \Lambda^{\rm top}V$,
and it can be shown \cite{dgks}[section 3.3] that this is generated by
\begin{displaymath}
\det\left( \psi \tilde{A} \: + \: \tilde{\psi} \tilde{B} \right),
\end{displaymath}
where $\{ \psi, \tilde{\psi} \}$ is a basis for $W$.
Thus, the kernel of $H^{n}(S_{n}) \rightarrow H^{n+1}(S_{n+1})$
is generated by
\begin{displaymath}
\det\left( \psi \tilde{A} \: + \: \tilde{\psi} \tilde{B} \right).
\end{displaymath}

The case $i-1=m\neq n$ is nearly identical, so we omit its description.
If $i-1=n=m$, the result is very similar.  In this case,
\begin{displaymath}
H^{n}(\Lambda^{n+1} Z) \: = \: \Lambda^{n+1} V \oplus \Lambda^{n+1} \tilde{V}
\end{displaymath}
and the kernel of $H^{n}(S_{n}) \rightarrow H^{n+1}(S_{n+1})$
is generated by
\begin{displaymath}
\det\left( \psi \tilde{A} \: + \: \tilde{\psi} \tilde{B} \right),
\: \: \:
\det\left( \psi \tilde{C} \: + \: \tilde{\psi} \tilde{D} \right).
\end{displaymath}

Putting this together, we find that the classical sheaf cohomology ring
of ${\bf P}^n \times {\bf P}^m$ with bundle ${\cal E}$ is 
generated by $\psi$, $\tilde{\psi}$ with relations
\begin{displaymath}
\det\left( \psi \tilde{A} \: + \: \tilde{\psi} \tilde{B} \right)  
\: = \: 0 \: = \:
\det\left( \psi \tilde{C} \: + \: \tilde{\psi} \tilde{D} \right).
\end{displaymath}

\subsection{Quantum sheaf cohomology ring on ${\bf P}^1 \times {\bf P}^1$}
\label{sect:qsc-p1p1}

Define
\begin{eqnarray*}
Q & \equiv & \det( \psi A \: + \: \tilde{\psi} B ), \\
\tilde{Q} & \equiv & \det( \psi C \: + \: \tilde{\psi} D ).
\end{eqnarray*}
In this section, we will show that
the quantum sheaf cohomology ring of ${\bf P}^1 \times {\bf P}^1$,
with bundle ${\cal E}$ defined earlier, is given by
\begin{displaymath}
{\bf C}[\psi, \tilde{\psi} ] / (Q - q, \tilde{Q} - \tilde{q} ).
\end{displaymath}
First, we shall derive the form of the cohomology ring in each
fixed instanton sector, then, we shall find relations between the sectors.

We shall begin by deriving the ring for 
fixed instanton degree $(d,e)$.  As outlined earlier,
the linear sigma model moduli space is computed to be
\begin{displaymath}
{\cal M} \: = \:
{\bf P}^{2d+1} \times {\bf P}^{2e+1}.
\end{displaymath}
Define $Z$ to be the following sheaf on ${\cal M}$:
\begin{displaymath}
Z \: \equiv \: \left(
{\rm Sym}^d U \otimes V \otimes {\cal O}(-1,0) \right) \oplus
\left(
{\rm Sym}^e U \otimes \tilde{V} \otimes {\cal O}(0,-1) \right).
\end{displaymath}
The induced sheaf ${\cal F}^*$ is the kernel 
\begin{displaymath}
0 \: \longrightarrow \: {\cal F}^* \: \longrightarrow \: Z \: 
\longrightarrow \: W \otimes {\cal O}_{\cal M} \: \longrightarrow \: 0,
\end{displaymath}
where
\begin{displaymath}
U \: = \: \Gamma({\bf P}^1, {\cal O}(1)), \: \: \:
V \: = \: \Gamma({\bf P}^1 \times {\bf P}^1, {\cal O}(1,0)), \: \: \:
\tilde{V} \: = \: \Gamma({\bf P}^1 \times {\bf P}^1,
{\cal O}(0,1)), \: \: \:
W \: = \: {\bf C}^2,
\end{displaymath}
which is naturally induced
from the short exact sequence defining ${\cal E}^*$, as discussed
in section~\ref{sect:genlprocedure}.

The desired correlation function in sector $(d,e)$ can be computed
as a classical sheaf cohomology cup product on 
${\cal M} = {\bf P}^{2d+1} \times {\bf P}^{2e+1}$.
As we have already computed classical sheaf cohomology on a product
of projective spaces, we can apply our results from the previous
subsection.  The induced maps are such that, for example,
\begin{displaymath}
\tilde{A} \: = \: {\rm diag}(A, A, \cdots, A)
\end{displaymath}
($d+1$ copies), hence the classical sheaf cohomology ring relations are
\begin{displaymath}
\det\left( \psi \tilde{A} \: + \: \tilde{\psi} \tilde{B} \right)
\: = \: \det\left( \psi A \: + \: \tilde{\psi} B \right)^{d+1} \: = \: 0,
\end{displaymath}
\begin{displaymath}
\det\left( \psi \tilde{C} \: + \: \tilde{\psi} \tilde{D} \right)
\: = \: \det\left( \psi C \: + \: \tilde{\psi} D \right)^{e+1} \: = \: 0,
\end{displaymath}
and so we immediately find that  
for fixed degree $(d,e)$, the sheaf cohomology
groups $$H^*\left({\cal M}, \Lambda^* {\cal F}^*\right)$$ 
live in the polynomial ring
\begin{displaymath}
{\rm Sym}^{\cdot} W \, / \, (Q^{d+1}, \tilde{Q}^{e+1}).
\end{displaymath}

For example, for degree $(d,e) = (1,0)$,
the kernel is spanned by the four polynomials
\begin{displaymath}
Q^2, \tilde{Q} \psi^2, \tilde{Q} \psi \tilde{\psi}, \tilde{Q} \tilde{\psi}^2
\end{displaymath}
and it is straightforward to check that this is a correct property of
the correlation functions give in {\it e.g.} \cite{ks}[equ'ns (21)-(30)].

It remains to derive the operator product ring, the quantum sheaf cohomology
ring utilizing the structure derived.

As there are no four-fermi contributions (${\cal F}_1 = {\rm Obs} = 0$), 
we expect from existence of
operator products that there should be relations between correlation functions
in different instanton sectors, of the form
\begin{equation}  \label{rreln}
\langle {\cal O} \rangle_{d,e} \: \propto \:
\langle {\cal O} R_{d,e,d',e'} \rangle_{d',e'}
\end{equation}
for all ${\cal O}$ and some fixed operator $R_{d,e,d',e'}$.  
For example,
\begin{displaymath}
\langle {\cal O} \rangle_{0,0} \: \propto \: \langle {\cal O} Q \rangle_{1,0},
\end{displaymath}
which suggests $Q = q$ for some proportionality constant $q$, and
\begin{displaymath}
\langle {\cal O} \rangle_{0,0} \: \propto \: \langle {\cal O}
\tilde{Q} \rangle_{0,1}
\end{displaymath}
which suggests $\tilde{Q} = \tilde{q}$ for some proportionality constant
$\tilde{q}$.  Equation~(\ref{rreln}) is merely the generalization to
arbitrary instanton degrees.
Because of compatibility with the
kernels above ({\it i.e.} maps must send kernels to (subsets of) kernels,
and must map top-forms to top-forms), the relations~(\ref{rreln})
should be of the form
\begin{displaymath}
\langle {\cal O} \rangle_{d,e} \: \propto \:
\langle {\cal O} Q^{d'-d} \tilde{Q}^{e'-e} \rangle_{d',e'}
\end{displaymath}
hence
\begin{displaymath}
\langle {\cal O} \rangle_{d,e} \: = \:
A_{d,e,d',e'} 
\langle {\cal O} Q^{d'-d} \tilde{Q}^{e'-e} \rangle_{d',e'}
\end{displaymath}
for some constant $A_{d,e,d',e'}$.
We assume that the constant $A_{d,e,d',e'}$ has the form
\begin{displaymath}
A_{d,e,d',e'} 
\: = \: q^{d'-d} \tilde{q}^{e'-e}
\end{displaymath}
for some constants $q$, $\tilde{q}$.
Note that mathematically this is an assumption, not a derivation; 
we justify this
assumption by the fact that this is the standard form of
nonperturbative corrections to operator products, and
so we recover standard physics results.

Thus,
\begin{displaymath}
\langle {\cal O} \rangle_{d,e} \: = \:
 q^{d'-d} \tilde{q}^{e'-e}
\langle {\cal O} Q^{d'-d} \tilde{Q}^{e'-e} \rangle_{d',e'},
\end{displaymath}
and in particular,
\begin{displaymath}
\langle \psi \tilde{\psi} Q^d \tilde{Q}^e \rangle_{d,e} \: = \:
q^d \tilde{q}^e \langle \psi \tilde{\psi} \rangle_{0,0},
\end{displaymath}
from which we derive the quantum sheaf cohomology relations
\begin{displaymath}
Q \: \sim \: q, \: \: \:
\tilde{Q} \: \sim \: \tilde{q},
\end{displaymath}
so that the quantum sheaf cohomology ring is given by
\begin{displaymath}
{\bf C}[\psi, \tilde{\psi} ] / (Q - q, \tilde{Q} - \tilde{q} ).
\end{displaymath}
This matches the prediction of \cite{mcom}, and also specializes to
the results in \cite{ks,gk}.

As a consistency check, let us quickly observe how bundle isomorphisms preserve the ring above.
Let $R \in GL(W)$, $P_1 \in GL(V)$, $P_2 \in GL(\tilde{V})$.
Under the action of
this $GL(2)^3$,
\begin{displaymath}
E \: \mapsto \: \left[ \begin{array}{cc}
P_1 A  x & P_1 B  x \\
P_2 C  \tilde{x} & P_2 D  \tilde{x} \end{array}
\right] R.
\end{displaymath}
As $R$ also acts on $\psi$, $\tilde{\psi}$,
its action falls out of the ring relations, and we are left with
\begin{eqnarray*}
\det\left( A \psi \: + \: B \tilde{\psi} \right) & \mapsto &
\det \left( P_1 \left( A \psi \: + \: B \tilde{\psi} \right) \right)
\: = \: \det P_1 \det\left( A \psi \: + \: B \tilde{\psi} \right), \\
\det \left( C \psi \: + \: D \tilde{\psi} \right) & \mapsto &
\det \left( P_2 \left( C \psi \: + \: D \tilde{\psi} \right) \right)
\: = \: \det P_2 \det \left( C \psi \: + \: D \tilde{\psi} \right),
\end{eqnarray*}
and so we see that by absorbing $\det P_i$ into $q$, $\tilde{q}$,
the ring is preserved.

\section{Example:  Hirzebruch surface}
\label{sect:ex:hirz}

Next, we shall compute quantum sheaf cohomology for a deformation
of the
tangent bundle of the Hirzebruch surface ${\bf F}_n$.
We will use the same notation as earlier in section~\ref{sect:genlprocedure}.
As in that section, 
the homogeneous coordinates $u$, $v$, $s$, $t$ (corresponding to the
four toric divisors) have the following weights under two ${\bf C}^{\times}$
actions:
\begin{center}
\begin{tabular}{cccc}
$u$ & $v$ & $s$ & $t$ \\ \hline
$1$ & $1$ & $0$ & $n$ \\
$0$ & $0$ & $1$ & $1$
\end{tabular}
\end{center}
We describe a deformation $\cE^*$ of the cotangent bundle as the
kernel
\begin{displaymath}
0 \: \longrightarrow \: \cE^* \: \longrightarrow \: Z
 \: \stackrel{E}{\longrightarrow} \: W \otimes
{\mathcal O} \: \longrightarrow \: 0,
\end{displaymath}
where 
\begin{displaymath}
Z \: = \: {\mathcal O}(-1,0)^{\oplus 2} \oplus {\mathcal O}(0,-1) \oplus
{\mathcal O}(-n,-1), 
\end{displaymath}
$W$ is a two-dimensional vector space,
\begin{displaymath}
E \: = \: \left[ \begin{array}{cc}
A x & B x \\
\gamma_1 s & \gamma_2 s \\
\alpha_1 t + s f_1(u,v) & \alpha_2 t + s f_2(u,v)
\end{array} \right],
\end{displaymath}
with
\begin{displaymath}
x \: \equiv \: \left[ \begin{array}{c} u \\ v \end{array} \right],
\end{displaymath}
$A$, $B$ constant $2 \times 2$ matrices, $\gamma_1$, $\gamma_2$,
$\alpha_1$, $\alpha_2$ constants, and $f_{1,2}(u,v)$ homogeneous polynomials
of degree $n$.

First, we shall outline the classical cohomology ring.
As before, we use the generalized
Koszul complex associated to $\Lambda^2 {\cal E}^*$,
split it into two short exact sequences, and compute the kernel of the
map ${\rm Sym}^2 W \rightarrow H^2(\Lambda^2 {\cal E}^*)$.
The kernel arises from $H^1(\Lambda^2 Z)$, which is two-dimensional.
It can be shown \cite{dgks}[section 3.3] that the kernel is
generated by
\begin{displaymath}
\det\left( \psi A \: + \: \tilde{\psi} B \right), \: \: \:
\left( \psi \gamma_1 \: + \: \tilde{\psi} \gamma_2 \right)
\left( \psi \alpha_1 \: + \: \tilde{\psi} \alpha_2 \right).
\end{displaymath}

Because we will be encountering these polynomials often, we shall assign
them names as follows:
\begin{eqnarray*}
Q_{K1} & = & \det \left( \psi A \: + \: \tilde{\psi} B \right), \\
Q_s & = &  \psi \gamma_1 \: + \: \tilde{\psi} \gamma_2, \\
Q_t & = & \psi \alpha_1 \: + \: \tilde{\psi} \alpha_2.
\end{eqnarray*}
(This nomenclature is used in the companion paper \cite{dgks}.)
Thus, the kernel in the degree $\vec{d}=0$ sector is generated by
$Q_{K1}$, $Q_s Q_t$.

Next, consider the sector of instanton degree $\vec{d}=(1,0)$.
The linear sigma model moduli space has homogeneous coordinates
$u_{0,1}$, $v_{0,1}$, $s$, $t_{0, \cdots, n}$, with weights
\begin{center}
\begin{tabular}{cccc} 
$u_{0,1}$ & $v_{0,1}$ & $s$ & $t_{0,\cdots,n}$ \\ \hline
$1$ & $1$ & $0$ & $n$ \\
$0$ & $0$ & $1$ & $1$ 
\end{tabular}
\end{center}
with exceptional set
\begin{displaymath}
\{ u_0 = u_1 = v_0 = v_1 = 0, \: \: \:
s = t_0 = t_1 = \cdots = t_n = 0 \}.
\end{displaymath}
The induced bundle ${\cal F}$ is given by
\begin{displaymath}
0 \: \longrightarrow \: {\cal F}^* \: \longrightarrow \: Z \: 
\stackrel{E'}{\longrightarrow} \: W \otimes {\cal O} \: 
\longrightarrow \: 0,
\end{displaymath}
where
\begin{displaymath}
Z \: = \: {\cal O}(-1,0)^{\oplus 4} \oplus {\cal O}(0,-1) \oplus
{\cal O}(-n,-1)^{\oplus \, n+1}
\end{displaymath}
and the map $E'$, induced from $E$, is given by
\begin{displaymath}
E' \: = \: \left[ \begin{array}{cc}
A x_0 & B x_0 \\
A x_1 & B x_1 \\
\gamma_1 s & \gamma_2 s \\
\alpha_1 t_0 \: + \: s f_1(u_0, v_0) & 
\alpha_2 t_0 \: + \: s f_2(u_0, v_0) \\
\alpha_1 t_1 \: + \: s \cdots & \alpha_2 t_1 \: + \: s \cdots \\
\cdots & \cdots \\
\alpha_1 t_n \: + \: s \cdots & \alpha_2 t_n \: + \: s \cdots
\end{array} \right],
\end{displaymath}
where we are using $\cdots$ to abbreviate full zero mode expansions
of $f_1$, $f_2$, as described earlier in {\it e.g.} 
the analogous case of equation~(\ref{fullzeromodeexp}).
We use $s \cdots$ merely to denote a series of terms which have $s$ as
a common factor.

We want to compute the kernel of the map ${\rm Sym}^{n+4}W \rightarrow
H^{n+4}(\Lambda^{n+4} {\cal F}^*)$, which we do using the generalized
Koszul complex associated
to $\Lambda^{n+4} {\cal F}^*$.  Following the usual pattern, and using
the exceptional set described above (and the primitive collection
it determines as in \cite{dgks}), we find that the map
$H^3(S_3) \rightarrow H^4(S_4)$ fails to be an isomorphism
(because $H^3(\Lambda^4 Z)$ is nonzero) and $H^{n+1}(S_{n+1}) \rightarrow
H^{n+2}(S_{n+2})$ fails to be an isomorphism (because
$H^{n+1}(\Lambda^{n+2} {\cal F})$ is nonzero).  The kernel arising from the
first is generated by \cite{dgks}[section 3.3]
\begin{displaymath}
Q_{K1}^2 \: = \: 
\left( \det \left( \psi A \: + \: \tilde{\psi} B \right) \right)^2
\end{displaymath}
and the kernel arising from the second is generated by \cite{dgks}[section 3.3]
\begin{displaymath}
Q_s Q_t^{n+1} \: = \: 
\left( \psi \gamma_1 \: + \: \tilde{\psi} \gamma_2 \right)
\left( \psi \alpha_1 \: + \: \tilde{\psi} \alpha_2 \right)^{n+1}.
\end{displaymath}

In terms of correlation functions, the result above implies that
\begin{displaymath}
\langle {\cal O} \rangle_{\vec{d} = 0} \: \propto \:
\langle {\cal O} Q_{K1} Q_t^n
\rangle_{\vec{d} = (1,0)},
\end{displaymath}
which suggests that the OPE ring has the (partial) form
\begin{equation}  \label{hirzreln1}
Q_{K1} Q_t^n \: = \: q_1
\end{equation}
for some parameter $q_1$.

Next, consider the degree $\vec{d} = (0,1)$ sector.
The linear sigma model moduli space has homogeneous coordinates
$u$, $v$, $s_{0,1}$, $t_{0,1}$ (where the $s_i$ and $t_i$ are
the coefficients in the zero-mode expansion of $s$, $t$).
These coordinates have weights:
\begin{center}
\begin{tabular}{cccccc}
$u$ & $v$ & $s_0$ & $s_1$ & $t_0$ & $t_1$ \\ \hline
$1$ & $1$ & $0$ & $0$ & $n$ & $n$ \\
$0$ & $0$ & $1$ & $1$ & $1$ & $1$
\end{tabular}
\end{center}
and the exceptional set is given by
\begin{displaymath}
\{ u=v=0, \: \: \: s_0 = s_1 = t_0 = t_1 = 0 \}.
\end{displaymath}
The induced bundle ${\cal F}$ is now given by
\begin{displaymath}
0 \: \longrightarrow \: {\cal F}^* \: \longrightarrow \:
Z \: \stackrel{E'}{\longrightarrow} \: W \otimes {\cal O} \:
\longrightarrow \: 0,
\end{displaymath}
where 
\begin{displaymath}
Z \: = \: {\cal O}(-1,0)^{\oplus 2} \oplus {\cal O}(0,-1)^{\oplus 2}
\oplus {\cal O}(-n,-1)^{\oplus 2},
\end{displaymath}
and the map $E'$, induced from $E$, is given by
\begin{displaymath}
E' \: = \: \left[ \begin{array}{cc}
A x & B x \\
\gamma_1 s_0 & \gamma_2 s_0 \\
\gamma_1 s_1 & \gamma_2 s_1 \\
\alpha_1 t_0 \: + \: s_0 f_1(u,v) & \alpha_2 t_0 \: + \: s_0 f_2(u,v) \\
\alpha_1 t_1 \: + \: s_1 f_1(u,v) & \alpha_2 t_1 \: + \: s_1 f_2(u,v)
\end{array} \right].
\end{displaymath}
As before, we want to compute the kernel of the map
${\rm Sym}^4 W \rightarrow H^4(\Lambda^4 {\cal F}^*)$,
which we do using the generalized Koszul complex associated to
$\Lambda^4 {\cal F}^*$.
Following the usual pattern, and using the exceptional collection
described above, we find that the map
$H^1(S_1) \rightarrow H^2(S_2)$ fails to be an isomorphism
(because $H^1(\Lambda^2 Z)$ is nonzero) and
$H^3(S_3) \rightarrow H^4(\Lambda^4 {\cal F}^*)$ fails to be an
isomorphism (because $H^3(\Lambda^4 Z)$ is nonzero).
The kernel arising from the first is generated by
\begin{displaymath}
Q_{K1} \: = \: \det\left( \psi A \: + \: \tilde{\psi} B \right)
\end{displaymath}
and the kernel arising from the second is generated by
\cite{dgks}[section 3.3]
\begin{displaymath}
Q_s^2 Q_t^2 \: = \: \left( \psi \gamma_1 \: + \: \tilde{\psi} \gamma_2 \right)^2
\left( \psi \alpha_1 \: + \: \tilde{\psi} \alpha_2 \right)^2.
\end{displaymath}

In terms of correlation functions, the result above implies that
\begin{displaymath}
\langle {\cal O} \rangle_{\vec{d} = 0} \: \propto \:
\langle {\cal O} Q_s Q_t
\rangle_{\vec{d} = 
(0,1) }
\end{displaymath}
which suggests that the OPE ring has the (partial) form
\begin{equation}  \label{hirzreln2}
Q_s Q_t \: = \: q_2
\end{equation}
for some parameter $q_2$.

Now, consider the sector of instanton degree
$\vec{d}=(1,-n)$.  In this sector, we need to take into account
contributions from four-fermi terms, something we have not needed
to do previously.
The linear sigma model moduli space has homogeneous coordinates
$u_{0,1}$, $v_{0,1}$, $t$ (where the $u_i$, $v_i$ are the coefficients
in the zero mode expansion of $u$, $v$, and $s$ does not contribute because
it has no zero modes in this sector).  These coordinates have weights
\begin{center}
\begin{tabular}{ccccc}
$u_0$ & $u_1$ & $v_0$ & $v_1$ & $t$ \\ \hline
$1$ & $1$ & $1$ & $1$ & $n$ \\
$0$ & $0$ & $0$ & $0$ & $1$
\end{tabular}
\end{center}
and the exceptional set is given by
\begin{displaymath}
\{ u_0 = u_1 = v_0 = v_1 = 0, \: \: \:
t = 0 \}.
\end{displaymath}
The induced bundle ${\cal F}$ is given by
\begin{displaymath}
0 \: \longrightarrow \: {\cal F}^* \: \longrightarrow \: Z \:
\stackrel{E'}{\longrightarrow} \: W \otimes {\cal O} \:
\longrightarrow \: 0,
\end{displaymath}
where
\begin{displaymath}
Z \: = \: {\cal O}(-1,0)^{\oplus 4} \oplus {\cal O}(-n,-1),
\end{displaymath}
and the map $E'$, induced from $E$, is given by
\begin{displaymath}
E' \: = \: \left[ \begin{array}{cc}
A x_0 & B x_0 \\
A x_1 & B x_1 \\
\alpha_1 t & \alpha_2 t
\end{array} \right],
\end{displaymath}
where $x_i = [u_i, v_i]^T$.
Furthermore, the second $U(1)$ effectively removes $t$ from the moduli
space, so the linear sigma model moduli space is effectively 
${\bf P}^3$, and then\footnote{
The reader might ask why the last factor is ${\cal O}$ instead of
${\cal O}(-n)$, since it arises from ${\cal O}(-n,-1)$.  The answer is that
the ${\bf C}^{\times}$ action describing the ${\bf P}^3$, must leave the
$t$ coordinate neutral.  If we label the two ${\bf C}^{\times}$ actions
defining ${\bf F}_n$ as $\lambda$, $\mu$, then the ${\bf C}^{\times}$
action defining ${\cal M} = {\bf P}^3$ is $\lambda - n \mu$, so that
over that ${\cal M}$, $t$ is a smooth section of ${\cal O}$ and
$s$ is a smooth section of ${\cal O}(-n)$.
}
$Z = {\cal O}(-1)^{\oplus 4} \oplus {\cal O}$.  
In this example, ${\cal F}_1$ will be nonzero, as we will discuss
momentarily, but first let us compute the cohomology ring structure
in this instanton sector.

Proceeding based on previous experience, the kernel will have
two components.
One component will arise from $H^3(\Lambda^4 Z) \neq 0$.
This kernel will be proportional to
\begin{displaymath}
Q_{K1}^2 \: = \:
\left( \det \left( \psi A \: + \: \tilde{\psi} B \right) \right)^2.
\end{displaymath}
The second component will arise from $H^0(Z) \neq 0$.
This kernel will be proportional to
\begin{displaymath}
Q_t \: = \: \alpha_1 \psi \: + \: \alpha_2 \tilde{\psi}.
\end{displaymath}

Now, let us compute ${\cal F}_1$.  We will find that four-fermi terms
will contribute, something that has not been true in previous cases.
(As a result, the interpretation of the kernels computed above as
kernels of correlation functions is more subtle than before -- in
some ways, this case is more closely parallel to the details of
a single projective
space and the kernels computed there.)
Here,
\begin{displaymath}
{\cal F}_1 \: = \: H^1\left( {\bf P}^1, {\cal O}(-n) \right) 
\otimes {\cal O}(0,1) \: = \: \oplus_1^{n-1} {\cal O}(0,1)
\end{displaymath}
(for $n \geq 1$; we omit $n=0$ as we have already studied
${\bf P}^1 \times {\bf P}^1$).  If we describe the moduli space as
${\bf P}^3$, then ${\cal F}_1 = {\cal O}(-n)^{\oplus n-1}$.
(In previous cases, ${\cal F}_1$ vanished; we only mention it when it
is nonzero.)

Since ${\cal F}_1$ is nonzero (and of the same rank as the obstruction
bundle, which in fact is identical), in each correlation function in this
sector we need to insert 
\begin{displaymath}
Q_s^{n-1} \: = \:
\left( \psi \gamma_1 \: + \: \tilde{\psi} \gamma_2 \right)^{n-1}
\end{displaymath}
(following appendix~\ref{sect:glsm-4fermi}).

Now, let us find some relations between correlation functions.
First, let us relate correlation functions in degree $(1,-n)$ to
those in degree $(1,0)$.  In both degrees, $Q_{K1}^2$ partially
generates the kernel, but in the former case, the rest arises from
$Q_t$, whereas in the latter case, $Q_s Q_t^{n+1}$ is a generator,
so to account for the difference, to map kernels to kernels,
correlators in the degree $(1,0)$ sector must be multiplied by
$Q_s Q_t^{n+1}/Q_t = Q_s Q_t^n$.  Furthermore, because in the
degree $(1,-n)$ sector, four-fermi terms add a factor of $Q_s^{n-1}$,
we must also add that same factor to correlators in degree $(1,0)$.
Thus, we find that
\begin{displaymath}
\langle {\cal O} \rangle_{\vec{d} = (1,-n)} \: \propto \:
\langle {\cal O} \left( Q_s Q_t^n \right) \left( Q_s^{n-1} \right)
\rangle_{\vec{d} = (1,0)} \: = \:
\langle {\cal O} \left( Q_s Q_t \right)^n \rangle_{\vec{d}=(1,0)}.
\end{displaymath}
Note that this result is compatible with the earlier relation~(\ref{hirzreln1}),
namely
\begin{displaymath}
Q_s Q_t \: = \: q_2
\end{displaymath}
for some constant $q_2$; furthermore, to achieve that compatibility
required both matching kernels and also utilizing four-fermi terms.

As one more consistency check, let us now work out the relation between
correlation functions in degree $\vec{d} = 0$ and those in
degree $\vec{d} = (1,-n)$.
In the former case, the kernel is generated by $Q_{K1}$, $Q_s Q_t$,
whereas in the latter case, the kernel is generated by $Q_{K1}^2$,
$Q_t$, so if we ignore four-fermi terms, then to match kernels,
correlation functions would be related by
\begin{displaymath}
\langle {\cal O} Q_s \rangle_{\vec{d}=0} \: \propto \:
\langle {\cal O} Q_{K1} \rangle_{\vec{d}=(1,-n)}.
\end{displaymath}
Because in degree $(1,-n)$ we also have four-fermi terms, generating
factors of $Q_s^{n-1}$, the correct relation between correlation functions
is
\begin{displaymath}
\langle {\cal O} Q_s Q_s^{n-1} \rangle_{\vec{d}=0} \: \propto \:
\langle {\cal O} Q_{K1} \rangle_{\vec{d}=(1,-n)}.
\end{displaymath}
In terms of our previous relations, this suggests that
\begin{displaymath}
Q_{K1}  \: = \: q_1 q_2^{-n} Q_s^n,
\end{displaymath}
which is indeed
an algebraic consequence of~(\ref{hirzreln1}), (\ref{hirzreln2}).
(Simply multiply both sides by either $Q_s^n$ or $Q_t^n$ and apply
(\ref{hirzreln2}) to turn one into the other.)
Again, note we need both kernels and four-fermi terms to derive
consistent relations.

This last example also illustrates a technical point regarding
OPE computations that will arise in \cite{dgks}.  There, we will derive
OPE's by giving relations between correlation functions of the form
\begin{displaymath}
\langle {\cal O} \rangle_{\beta} \: \propto \:
\langle {\cal O} R_{\beta, \beta'} \rangle_{\beta'}.
\end{displaymath}
Except for the last case above, in the examples we have studied
it has been possible to find an $R_{\beta,\beta'}$
putting relations in the form above.  However, the last example illustrates
that this cannot always be done.  Technically, in \cite{dgks} we deal
with this issue through the introduction of `direct systems' to describe
relations between correlation functions of different degrees.

Let us also take a moment to discuss the interpretation of the $q$'s.
In this text, we have been using them merely as placeholders for unspecified
constants; in particular, classical limits do not necessarily correspond
to the case that all $q_i \rightarrow 0$.  To clarify this, let us consider
the (2,2) limit of the relations we have been deriving.
In this limit,
\begin{displaymath}
A \: = \: I, \: \: \: B \: = \: 0, \: \: \:
\gamma_1 \: = \: 0, \: \: \: \gamma_2 \: = \: 1, \: \: \:
\alpha_1 \: = \: n, \: \: \: \alpha_2 \: = \: 1, \: \: \:
f_1 \: = \: f_2 \: = \: 0
\end{displaymath}
As a result,
\begin{displaymath}
Q_{K1} \: = \: \psi^2, \: \: \:
Q_s \: = \: \tilde{\psi}, \: \: \: Q_t \: = \: n \psi \: + \: \tilde{\psi}
\end{displaymath}
The classical cohomology ring of the Hirzebruch surface can be described
by (toric) generators $D_u$, $D_v$, $D_s$, $D_t$ in degree 2, obeying
\begin{displaymath}
D_u \: \sim D_v, \: \: \: D_t \: \sim \: D_s \: + \: n D_v
\end{displaymath}
\begin{displaymath}
D_u^2 \: = \: 0, \: \: \: D_s( n D_u \: + \: D_s) \: = \: 0
\end{displaymath}
If we identify $D_u = \psi$, $D_s = \tilde{\psi}$, then the 
relations~(\ref{hirzreln1}), (\ref{hirzreln2}), namely,
\begin{displaymath}
Q_{K1} Q_t^n \: = \: q_1, \: \: \: Q_s Q_t \: = \: q_2
\end{displaymath}
become
\begin{displaymath}
D_u^2 (n D_t \: + \: D_u) \: = \: q_1, \: \: \:
D_s( n D_u \: + \: D_s) \: = \: q_2
\end{displaymath}
which clearly do not have the correct classical limit when $q_1 \rightarrow
0$.  On the other hand, the equivalent relations
\begin{displaymath}
Q_{K1} \: = \: q' Q_s^n, \: \: \:
Q_s Q_t \: = \: q_2
\end{displaymath}
(where $q' = q_1 q_2^{-n}$) become
\begin{displaymath}
D_u^2 \: = \: q' D_s^n, \: \: \: 
D_s( n D_u \: + \: D_s ) \: = \: q_2
\end{displaymath}
which does reduce to the classical cohomology ring relations
when $q' = q_1 q_2^{-n}
 \rightarrow 0$ and
$q_2 \rightarrow 0$.  This also illuminates the issue with the previos
presentation -- $q_1 \rightarrow 0$, $q_2 \rightarrow 0$ independently do not
give the classical limit, one must also demand $q_1 q_2^{-n} \rightarrow 0$.
Thus, we see that only certain presentations of the 
ring will give the classical cohomology ring on the (2,2) locus when
all $q \rightarrow 0$.  For other presentations, more complicated
limits must be taken\footnote{
We would like to thank I.~Melnikov for illuminating discussions of
this point.
}.  The particular presentation given in \cite{dgks} does have
the property that on the (2,2) locus, one recovers the correct classical
limit as all $q \rightarrow 0$ independently.
In this paper we shall not belabor this point.

Now, let us summarize.  We have not described an exhaustive survey
of all possibilities (see instead \cite{dgks}), but based on the
computations performed, it would seem that the OPE ring in this
example is defined by
\begin{eqnarray*}
Q_{K1} Q_t^n & = & q_1, \\
Q_s Q_t & = & q_2,
\end{eqnarray*}
which are relations~(\ref{hirzreln1}), (\ref{hirzreln2}).
We will see in section~\ref{sect:genl} that this is a correct
specialization of the general results of \cite{dgks}.

\section{General result}
\label{sect:genl}

\subsection{Result}

First, we shall outline the result from \cite{dgks}, and then compute it
in examples.

Let $\{\rho_i\}$ denote the (one-dimensional)
edges of the fan, {\it i.e.} the toric divisors,
and let $K_i$ denote
`primitive collections' of edges, that is, maximal 
collections of edges not contained in
any single cone.  (These collections define the Stanley-Reisner ideal,
through the statement that the toric divisors do not all intersect.)

To each primitive collection $K$, we can associate a unique divisor class
$\beta_K$, as follows.  Let the 
vector generating the edge of the fan corresponding
to $\rho$ be denoted $v_{\rho}$, then for $K = \{ \rho_1, \cdots, \rho_k \}$,
we can write
\begin{equation}    \label{cdefn}
v_{\rho_1} \: + \: \cdots \: + \: v_{\rho_k} \: = \:
\sum_{\rho} c_{\rho} v_{\rho}
\end{equation}
for some integers $c_{\rho} > 0$, with the sum on the right running over
toric divisors not necessarily in $K$.  
By moving the right-hand-side to the left, we can
write this as
\begin{equation}   \label{adefn}
\sum_{\rho} a_{\rho} v_{\rho} \: = \: 0
\end{equation}
for some integers $a_i$.  Then, it can be shown \cite{batyrev} that there
is a unique curve class $\beta_K$ such that $\beta_K \cdot \rho = a_{\rho}$
for all $\rho$.

Now, for each divisor class $c = [\rho]$, we define a $|c|\times|c|$ matrix
$A_c$, where $|c|$ is the number of toric divisors linearly equivalent
to $\rho$.  The matrix $A_c$ is given by the rows of the map $E$ appearing
in the definition of the deformation ${\cal E}^*$, the rows corresponding
to representatives of $c$, and with nonlinear terms omitted.
Define
\begin{displaymath}
Q_c \: = \: \det A_c.
\end{displaymath}

The quantum sheaf cohomology ring is then given by polynomials in the
elements of a basis for $W$, modulo the relations
\begin{equation}   \label{qsc-general}
\prod_{c \in [K]} Q_c \: = \: q^{\beta_K} \prod_{c \in [K^-]}
Q_c^{-d_c^{\beta_K}}
\end{equation}
for each primitive collection $K$, where $[K^-]$ denotes
the set of linear equivalence classes of edges appearing in the
right-hand-side of~(\ref{cdefn}) with nonzero coefficients $c_{\rho}$,
and $d_c^{\beta_K} \equiv c \cdot \beta_K$. 
(Note that for $c \in [K^-]$, the exponent $-d_c^{\beta_K}$ is 
nonnegative.)

The formula above gives a canonical presentation of the quantum sheaf
cohomology ring for each toric variety, dependent only upon the bundle and toric
variety and not the details of any particular presentation such
as ${\bf C}^{\times}$ weights or $U(1)$ charges in a quotient.

Let us work through a few examples of this formalism,
beginning with a projective space ${\bf P}^n$, as described in
section~\ref{sect:oneproj}.  Here, there are $n+1$ toric divisors
$\rho_0, \cdots, \rho_{n}$.  There is only one primitive collection,
\begin{displaymath}
K \: = \: \{ \rho_0, \cdots, \rho_n \},
\end{displaymath}
and for any fan, the vectors generating the
edges obey
\begin{displaymath}
v_{\rho_0} \: + \: \cdots \: + \: v_{\rho_n} \: = \: 0.
\end{displaymath}
The unique divisor class $\beta$ such that $\beta \cdot \rho = 1$ for
all $\rho$ is represented by any of the toric divisors $\rho$, since
they are all linearly equivalent.
All of the toric divisors are linearly equivalent, and so there is one
matrix $A_c = A \psi$, derived from the map $E$ defining ${\cal E}^*$,
and one $Q = \det A_c = \det (A \psi)$.
The quantum sheaf cohomology ring is then ${\bf C}[\psi]$ modulo the
relation
\begin{displaymath}
\det( A \psi) \: = \: q
\end{displaymath}
matching what was found in section~\ref{sect:oneproj}, and for that
matter matching the (2,2) locus (since on a single projective space,
all toric Euler deformations return the tangent bundle itself).

A slightly more interesting example is ${\bf P}^1 \times {\bf P}^1$,
as discussed in section~\ref{sect:ex:proj}.
Here, let $D_{x_{0,1}}$, $D_{\tilde{x}_{0,1}}$ denote the four toric divisors.
(We are using here the nearly same notation for the toric divisors that we used
for corresponding homogeneous coordinates in section~\ref{sect:ex:proj}.)
There are two primitive collections:
\begin{displaymath}
K_1 \: = \: \{ D_{x_0}, D_{x_1} \}, \: \: \:
K_2 \: = \: \{ D_{\tilde{x}_0}, D_{\tilde{x}_1} \}.
\end{displaymath}
In each primitive collection, the constituent divisors are all linearly
equivalent to one another.  Moreover, 
\begin{displaymath}
v_{x_0} \: + \: v_{x_1} \: = \: 0, \: \: \:
v_{\tilde{x}_0} \: + \: v_{\tilde{x}_1} \: = \: 0.
\end{displaymath}  
It is easy to check that $\beta_1$ is represented by $D_{\tilde{x}_0}$, 
$D_{\tilde{x}_1}$,
and $\beta_2$ is represented by $D_{x_0}$, $D_{x_1}$,
since $D_{x_i} \cdot D_{\tilde{x}_j} = 1$.
Following the notation of section~\ref{sect:ex:proj},
we find
\begin{eqnarray*}
A_1 \: = \: \psi A \: + \: \tilde{\psi} B, \\
A_2 \: = \: \psi C \: + \: \tilde{\psi} D,
\end{eqnarray*}
and from~(\ref{qsc-general}) the quantum sheaf cohomology ring 
is ${\bf C}[\psi,\tilde{\psi}]$ modulo the relations
\begin{eqnarray*}
\det\left( \psi A \: + \: \tilde{\psi} B \right) \: = \: q^{\beta_1}, \\
\det\left( \psi C \: + \: \tilde{\psi} D \right) \: = \: q^{\beta_2},
\end{eqnarray*}
matching the results of section~\ref{sect:ex:proj}.

Now, let us specialize the general result to Hirzebruch surfaces, and
compare to the results we obtained previously.
To that end, we describe the Hirzebruch
surface classically with four (toric) divisors
$D_u$, $D_v$, $D_s$, $D_t$, where
\begin{displaymath}
D_u \: = \: D_v, \: \: \:
D_t \: = \: D_s \: + \: n D_v
\end{displaymath}
and 
\begin{displaymath}
D_u \cdot D_v \: = \: 0 \: = \:
D_s \cdot D_t.
\end{displaymath}
There are two `primitive collections' of divisors, defined by the
Stanley-Reisner ideal above:
\begin{displaymath}
K_1 \: = \: \{ D_u, D_v \}, \: \: \:
K_2 \: = \: \{ D_s, D_t \}.
\end{displaymath}
For the first primitive collection,
\begin{displaymath}
v_u \: + \: v_v \: = \: n v_s,
\end{displaymath}
and the unique divisor class $\beta_{1}$ such that
\begin{displaymath}
[D_u] \cdot \beta_{1} \: = \: 1 \: = \: [D_v] \cdot \beta_{1}, \: \: \:
[D_s] \cdot \beta_{1} \: = \: - n, \: \: \:
[D_t] \cdot \beta_{1} \: = \: 0
\end{displaymath}
is represented by $D_s$, {\it i.e.} $\beta_{1} = [D_s]$.
Similarly,
\begin{displaymath}
v_s \: + \: v_t \: = \: 0
\end{displaymath}
and the unique divisor class $\beta_{2}$ such that
\begin{displaymath}
[D_u] \cdot \beta_{2} \: = \: 0 \: = \: [D_v] \cdot \beta_{2}, \: \: \:
[D_s] \cdot \beta_{2} \: = \: 1 \: = \: [D_t] \cdot \beta_{2}
\end{displaymath}
is represented by $D_u$, $D_v$, {\it i.e.}
$\beta_{K2} = [D_u] = [D_v]$.

Then, to each primitive collection $K$ is associated a polynomial in
the generators of $W$.  In this case, these polynomials are:
\begin{eqnarray*}
\prod_{c \in [K_1]} Q_c & = & Q_{K1}, \\
\prod_{c \in [K_2]} Q_c & = & Q_{s} Q_t.
\end{eqnarray*}
In the expressions above, note that $D_u$ and $D_v$ are linearly equivalent,
so there is only one linear equivalence class in $[K_1]$, but
$D_s$ and $D_t$ are not linearly equivalent, so there are two linear equivalence
classes in $[K_2]$.
Then, putting this together,
the quantum sheaf cohomology relations in~(\ref{qsc-general})
are
\begin{eqnarray*}
\prod_{c \in [K_1]} Q_c & = & Q_{K1} \: = \: 
q^{\beta 1} Q_{s}^{-d^{\beta_1}_s} \\
& = & q^{\beta_1} Q_{s}^{n}, \\
\prod_{c \in [K_2]} Q_c & = & Q_{s} Q_t \: = \: q^{\beta_2},
\end{eqnarray*}
where
\begin{displaymath}
d^{\beta_K}_{\rho} \: \equiv \: [D_{\rho}] \cdot \beta_K
\end{displaymath}
(here, $d^{\beta_1}_s = [D_s] \cdot \beta_{1} = -n$)
and $q^{\beta_1}$, $q^{\beta_2}$ are the two quantum parameters.

Note that by multiplying both sides by $Q_t^n$ and using the second
relation, we can turn these two relations into
\begin{eqnarray*}
Q_{K1} Q_t^n & = & q', \\
Q_s Q_t & = & q^{\beta_2},
\end{eqnarray*}
where $q' = q^{\beta_1}( q^{\beta}_2 )^n$.  It is this latter form in which 
the OPE rings for the Hirzebruch surface appear earlier in 
section~\ref{sect:ex:hirz}, where $q_1 = q'$, $q_2 = q^{\beta_2}$.

\subsection{Comparison to McOrist-Melnikov's results}

Let us now compare to the one-loop Coulomb branch results for the
quantum sheaf cohomology ring given in \cite{mcom}.  

Implicitly, the relations derived in all one-loop Coulomb branch computations
are not the relations of the ring at any single large-radius limit of the
GLSM, but rather live in a `localization' of the ring, in which operators
have been inverted.  Physically, this arises because the Coulomb branch
computations take place in a regime where $\sigma$ vevs are large, and so
can be assumed nonzero and invertible; mathematically, this makes it
possible for the one-loop Coulomb branch relations to be equally
applicable to all large-radius phases.  Thus, to compare the
results of the last subsection, derived in a single large-radius phase,
we must descend to a localization of the ring in which operator
invertibility is allowed, and make the comparison in that localization.
We will find that, after implicitly descending to that localization,
the results of the last subsection do indeed match the predictions of
\cite{mcom}.  

Partition the line bundle factors into collections
$\left\{ {\cal O}\left( \vec{q}_i \right)
\right\}$ with matching $\vec{q}_i$.
(We can think of this equivalently as partitioning the 
chiral superfields, indexed by $i$, into collections consisting of matching
$U(1)$ charges $\vec{q}_i$.)
Index such collections by $\alpha$.
(There is a one-to-one correspondence between such collections and
linear equivalence classes of toric divisors.)
Let
\begin{displaymath}
E_i: \:
{\cal O}^{\oplus r} \: \longrightarrow \:
{\cal O}\left(\vec{q}_i\right)
\end{displaymath}
denote the maps in the short exact sequence defining ${\cal E}$.
Define
\begin{displaymath}
A_{(\alpha)i}^{j a} \: \equiv \:
\left.
\frac{\partial}{\partial \phi_j} E_i^a \right|_{\phi \equiv 0}
\end{displaymath}
for $i$, $j$ in the collection $\alpha$.
For example, the tangent bundle of a toric variety is described by
\begin{displaymath}
E_i^a \: = \: Q_i^a \phi_i 
\end{displaymath}
hence
\begin{displaymath}
A_{(\alpha)i}^{j a} \: = \: \delta^j_i Q_{(\alpha)}^a,
\end{displaymath}
where $\vec{q}_\alpha = (Q^a_{\alpha})$ denotes the $U(1)$ charges
of all fields in the collection $\alpha$.

In this language, if we define $V_{\alpha}$ to be a vector space of the
same dimension as the number of line bundles in the collection $\alpha$
(the number of chiral superfields with matching charges $\vec{q}_{\alpha}$),
and let $W = {\bf C}^r$,
then we can describe the deformation of the tangent bundle as the
cokernel
\begin{displaymath}
0 \: \longrightarrow \: W \otimes {\cal O} \: \longrightarrow \:
\bigoplus_{\alpha} V_{\alpha} \otimes {\cal O}\left( \vec{q}_{\alpha} \right)
\: \longrightarrow \: {\cal E} \: \longrightarrow \: 0.
\end{displaymath}

Define
\begin{displaymath}
M_{(\alpha) i}^j \: = \: A_{(\alpha) i}^{ja} \psi_a.
\end{displaymath}
This is the same matrix that was denoted $A_c$ in the previous section,
but we have adapted our notation to more closely resemble that of
\cite{mcom}. 
In this notation, the result of \cite{dgks} is that the quantum sheaf cohomology
ring relations descend to 
\begin{equation}  \label{mm-general}
\prod_{\alpha} \left( \det M_{(\alpha)} \right)^{Q^a_{\alpha}} \: = \: q_{a}
\end{equation}
for each $a$, where $\vec{q}_{\alpha} = (Q^a_{\alpha})$,
and $q_a$ is the quantum parameter, modulo inversion of operators.

The ring above is specified in terms of the $U(1)$ charges of the
toric homogeneous coordinates, whereas in the previous section we gave
a canonical representation that was independent
of such choices.
Specifically, the canonical representative was described in terms
of $a_{\rho}$ defined in~(\ref{adefn}).  However, the charges $Q^a_{\alpha}$
are also defined as the kernel of a matrix formed from the $v_{\rho}$'s,
as in equation~(\ref{adefn}); thus, the $a_{\rho} = D_{\rho} \cdot \beta$ 
defined there are
precisely one set of charges.
With that in mind, the quantum sheaf cohomology relations~(\ref{qsc-general})
can be written in the form
\begin{displaymath}
\prod_c Q_c^{D_c \cdot \beta} \: = \: q^{\beta}
\end{displaymath}
for the $\beta$ associated to each primitive collection,
which is the same as
\begin{displaymath}
\prod_c \left( \det A_c \right)^{Q_c^{\beta}} \: = \: q^{\beta}
\end{displaymath}
for $Q_c^{\beta} \equiv D_c \cdot \beta$.
Thus, we see that the relations~(\ref{qsc-general}) specified in
\cite{dgks} really do descend to the relations~(\ref{mm-general})
in a localization\footnote{
We are implicitly performing this comparison in the localization
mentioned earlier, as we have not specified whether the charges $Q^a_{\alpha}$
are positive or negative.
} of the ring,
written in a form closer to that of reference~\cite{mcom}, for a
particular choice of charges $Q^a_{\alpha}$.

We should emphasize again that
this result is independent of nonlinear deformations
(meaning, terms in $E_i^a$ nonlinear in $\phi$'s),
as conjectured in {\it e.g.} \cite{mm2}[section 3.5].
This result also nicely meshes with previous physics results.
For example, \cite{kmmp}[section A.3] conjectured that A/2 correlation
functions should be independent of nonlinear deformations, based
on the fact that the discriminant locus in gauged linear
sigma models does not depend on such nonlinear deformations.

In the special case of linear deformations, {\it i.e.} when 
\begin{displaymath}
E_i^a \: = \: \sum_j A_{ (\alpha) i}^{j a} \phi_j,
\end{displaymath}
the result above specializes to the result of \cite{mcom},
computed with Coulomb branch techniques in gauged linear sigma models.

Now, let us compare to particular examples discussed earlier in this
paper.

In the case of deformations of the tangent bundle of
${\bf P}^1 \times {\bf P}^1$ discussed
in section~\ref{sect:qsc-p1p1}, it is
straightforward to check that there are two $M_{(\alpha)}$, given by
\begin{eqnarray*}
M_{(1)} & = & \psi_1 A \: + \: \psi_2 B, \\
M_{(2)} & = & \psi_1 C \: + \: \psi_2 D,
\end{eqnarray*}
and so we have the relations
\begin{displaymath}
\det M_{(1)} \: = \: q_1, \: \: \:
\det M_{(2)} \: = \: q_2,
\end{displaymath}
which matches our previous computation.

Next, let us describe the example of a Hirzebruch surface ${\bf F}_n$.
Consider a fan with edges $(1,0)$, $(0,1)$, $(-1,n)$, $(0,-1)$,
defined by the charges $(1,0)$, $(0,1)$, $(1,0)$, $(n,1)$
and homogeneous coordinates $u$, $s$, $v$, $t$, respectively,
as in figure~\ref{fig:FnFan}.
For a deformation of the tangent bundle of
${\bf F}_n$ as defined in~(\ref{hirz-genl-map}), we compute
\begin{eqnarray*}
M_{(1)} & = & A \psi_1 \: + \: B \psi_2, \\
M_{(2)} & = & \gamma_1 \psi_1 \: + \: \gamma_2 \psi_2, \\
M_{(3)} & = & \alpha_1 \psi_1 \: + \: \alpha_2 \psi_2,
\end{eqnarray*}
and so we have the quantum sheaf cohomology relations
\begin{eqnarray*}
\left( \det M_{(1)} \right) \left(  M_{(3)} \right)^n
 & = & q_1, \\
\left( M_{(2)} \right) \left( M_{(3)} \right)
 & = & q_2,
\end{eqnarray*}
which are precisely (\ref{hirzreln1}), (\ref{hirzreln2}) computed
earlier, identifying
\begin{displaymath}
Q_{K1} \: = \: \det M_{(1)}, \: \: \:
Q_s \: = \: M_{(2)}, \: \: \:
Q_t \: = \: M_{(3)}.
\end{displaymath}

In the special case that ${\cal E} = TX$, the ring above reduces to
\begin{displaymath}
\prod_i \left( \sum_b Q_i^b \psi_b \right)^{Q^a_i} \: = \: q_a,
\end{displaymath}
or equivalently,
\begin{eqnarray*}
\psi_1^2 \left( n \psi_1 \: + \: \psi_2 \right)^n & = & q_1, \\
\psi_2 \left( n \psi_1 \: + \: \psi_2 \right) & = & q_2,
\end{eqnarray*}
which is a standard result in (2,2) GLSM's \cite{dr}[equ'n (3.44)].
If we identify the toric divisors $D_i$ as
\begin{displaymath}
D_i \: = \: \sum_a Q_i^a \psi_a
\end{displaymath}
(as a consistency check, note that since $\sum_i Q_i^a \vec{v}_i = 0$,
it is necessarily the case that
\begin{displaymath}
\sum_i \langle m, \vec{v}_i \rangle D_i 
\: = \: 0, 
\end{displaymath}
hence the description above encodes the linear relations on
the Chow ring)
then the GLSM ring can be written as
\begin{displaymath}
\prod_i D_i^{Q_i^a} \: = \: q_a,
\end{displaymath}
or equivalently,
\begin{eqnarray*}
D_u D_v D_t^n & = & q_1, \\
D_s D_t & = & q_2,
\end{eqnarray*}
where 
\begin{displaymath}
D_u \: \sim \: D_v, \: \: \:
D_t \: \sim \: D_s \: + \: n D_v.
\end{displaymath}

There is an issue with classical limits, previously noted in
section~\ref{sect:ex:hirz}.  Let us outline the analysis here, in the
language of GLSM's.
The classical cohomology ring of ${\bf F}_n$ can be described by the
relations
\begin{displaymath}
D_u^2 \: = \: 0, \: \: \: 
D_s^2 \: = \: -n D_u D_s,
\end{displaymath}
and if we identify $D_u = \psi_1$, $D_s = \psi_2$, then we almost recover
this in the limit $q_a \rightarrow 0$, except for an extra factor
of $D_t^n$ modifying the relation $D_u^2=0$.
In order to make the relation with the classical cohomology ring more
clear, we should work in a different basis, one in which the fields
have charges $(1,0)$, $(1,0)$, $(0,1)$, $(-n,1)$.  (Note this is achieved
by an $SL(2,{\bf Z})$ transformation.)  In this basis, the quantum
cohomology ring becomes
\begin{displaymath}
\psi_1^2 \: = \: q_1 \left( -n \psi_1 \: + \: \psi_2 \right)^n, \: \: \:
\psi_2\left( -n \psi_1 \: + \: \psi_2 \right) \: = \: q_2
\end{displaymath}
and so when we set $q_a \rightarrow 0$, and identify
$D_u = \psi_1$, $D_s = - \psi_2$, we recover the classical cohomology ring
without extraneous factors.  (Alternatively, the canonical presentation of
the previous section avoids this problem.)
More invariantly, to cleanly recover the classical
cohomology relations from the GLSM relations, 
one wants to work in a basis such that the smooth phase
of the GLSM corresponds to the positive orthant of the secondary fan;
this is a property of the canonical presentation of the previous section.

\section{Conclusions}

In this paper we have outlined the mathematical computation of
quantum sheaf cohomology rings for deformations of tangent bundles of
toric varieties, emphasizing physics aspects of the computation.
Our new methods allow for much more efficient mathematical computations
than possible previously.  We have also seen in examples
that in these cases
(toric varieties, deformations of the tangent bundle), quantum sheaf
cohomology is independent of nonlinear deformations, as conjectured
elsewhere (see {\it e.g.} \cite{mm2,kmmp}).
Rigorous general proofs will appear in \cite{dgks}.

Extensions of the results of this paper
to Grassmannians and flag manifolds are under discussion
\cite{dgksnext}.  
Extensions to hypersurfaces would also be extremely useful.
In this paper and \cite{dgks} we compute kernels of correlation functions
in order to compute operator products;
it would also be interesting to work out complete expressions for
the correlation functions themselves.

\section{Acknowledgements}

We would like to thank M.~Ballard, J.~McOrist, and I.~Melnikov 
for useful conversations.
This work was presented in part at the conferences `String-Math 2011'
at the University of Pennsylvania, and `(0,2) mirror symmetry and
heterotic topological strings' at the Erwin Schr\"doinger Institute
in Vienna, Austria, and we thank both institutions for hospitality
while part of this work was completed.
R.D. was partially supported by NSF grants DMS-0908487, DMS-0636606.
J.G. was partially supported by NSF grant DMS-0636606.
S.K. was partially supported by NSF grant DMS-0555678.
E.S. was partially supported by NSF grants DMS-0705381, PHY-0755614.

\appendix

\section{GLSM derivation of four-fermi terms}   \label{sect:glsm-4fermi}

In this section we will outline how four-fermi effects arise in
gauged linear sigma models (GLSM's), and use this to derive the
ansatz for their effects in (0,2) theories used earlier in
sections~\ref{sect:ex:hirz}, \ref{sect:genl}.
Furthermore, we will see explicitly that
nonlinear deformations can not contribute to the four-fermi terms,
at least in the GLSM.

First, let us consider ordinary (2,2) supersymmetric gauged linear
sigma models for toric varieties, as in \cite{dr,edphases}.
As discussed in \cite{gs1}, correlation functions in the A-twisted
theory are correlation functions of products of $\overline{\sigma}$'s.
The GLSM itself does not contain any four-fermi terms, but the
effects of four-fermi terms in the low-energy nonlinear sigma model
are duplicated by Yukawa couplings in the GLSM of the form
\begin{displaymath}
\sum_{i,a} Q_i^a \overline{\sigma}_a \psi_z^{\overline{\imath}}
\psi_{\overline{z}}^i.
\end{displaymath}
Four-fermi terms in a nonlinear sigma model must be invoked whenever
$\psi_z^{\overline{\imath}}$, $\psi_{\overline{z}}^i$ have zero modes.
In the present case, when such fields have zero modes, to absorb them
one must use the Yukawa couplings above, which will then be
responsible for a factor of
\begin{displaymath}
\prod_i \left( \sum_a Q_i^a \overline{\sigma}_a \right)^{n_i},
\end{displaymath}
where
\begin{displaymath}
n_i \: = \: h^1\left( {\bf P}^1, {\cal O}\left( \vec{Q}_i \cdot \vec{d}
\right) \right),
\end{displaymath}
where $\vec{Q}_i$ has components $Q_i^a$, and $\vec{d}$ defines the
degree of the instanton sector.
(Strictly speaking, four-fermi interactions contribute integrals over of the
worldsheet of such factors; however, in the A model, correlators are
independent of position, so such integrals merely contribute factors of the
worldsheet area, which are cancelled out by corresponding factors in the
path integral.  See \cite{gs1} for details.)
This precisely duplicates the contribution described in
\cite{ks}[section 6.2.2],
\cite{dr}[equ'n (3.69)].

In an A/2 theory describing a toric variety with a deformation
of the tangent bundle, the results above are modified slightly.
There, the pertinent Yukawa couplings are of the form \cite{gs2,edphases}
\begin{displaymath}
\sum_{i,j} \psi_z^{\overline{\imath}} \psi_{\overline{z}}^j
\partial_{\overline{\imath}} \overline{E}^{\overline{\jmath}}.
\end{displaymath}

In terms of correlation functions, this means we insert
\begin{displaymath}
\prod_c \left( \det A_c \right)^{n_c},
\end{displaymath}
where $c$ runs over classes of toric divisors with the same
GLSM charge ({\it i.e.} linear equivalence classes), 
\begin{displaymath}
n_c \: = \: h^1\left({\bf P}^1, {\cal O}(\vec{Q}_c \cdot \vec{d}) \right),
\end{displaymath}
where $\vec{Q}_c$ is the charge of the homogeneous coordinates in
class $c$, and the matrix
\begin{displaymath}
A_c \: = \: \left( \partial_i E^j \right)
\end{displaymath}
with $i$, $j$ running in the same class $c$.
(Strictly speaking, again, four-fermi terms contribute integrals.
In the A/2 model, {\it a priori} worldsheet correlators are holomorphic
functions of position, but as argued in \cite{ade} for CFT's and
\cite{ilarion-priv} for GLSM's, in a neighborhood of
the (2,2) locus, A/2 correlators are actually independent of position, and so
again the integrals merely contribute factors of worldsheet area.)

Let us work out an example in detail to illustrate what this means.
Consider the example of a Hirzebruch surface ${\bf F}_n$, as in
section~\ref{sect:genlprocedure}.
In an example there, the deformation ${\cal E}$ of the tangent bundle
is described by
\begin{displaymath}
0 \: \longrightarrow \: {\cal E}^* \: \longrightarrow \:
{\cal O}(-1,0)^2 \oplus {\cal O}(0,-1) \oplus {\cal O}(-n,-1) \:
\stackrel{E}{\longrightarrow} \: W \otimes {\cal O} \:
\longrightarrow \: 0,
\end{displaymath}
where
\begin{displaymath}
E \: = \: \left[ \begin{array}{cc}
A x & B x \\
\gamma_1 s & \gamma_2 s \\
\alpha_1 t + s f_1(u,v) & \alpha_2 t + s f_2(u,v)
\end{array} \right],
\end{displaymath}
with
\begin{displaymath}
x \: \equiv \: \left[ \begin{array}{c} u \\ v \end{array} \right],
\end{displaymath}
$A$, $B$ constant $2 \times 2$ matrices, $\gamma_1$, $\gamma_2$,
$\alpha_1$, $\alpha_2$ constants, and $f_{1,2}(u,v)$ homogeneous polynomials
of degree $n$.

In the case above,
\begin{eqnarray*}
E^u & = & \left( A_{11} u \: + \: A_{12} v \right) \sigma_1 \: + \:
\left( B_{11} u \: + \: B_{12} v \right) \sigma_2, \\
E^v & = & \left( A_{21} u \: + \: A_{22} v \right) \sigma_1 \: + \:
\left( B_{21} u \: + \: B_{22} v \right) \sigma_2, \\
E^s & = & \gamma_1 s \sigma_1 \: + \: \gamma_2 s \sigma_2, \\
E^t & = & \left( \alpha_1 t \: + \: s f_1(u,v) \right) \sigma_1 \: + \:
\left( \alpha_2 t \: + \: s f_2(u,v) \right) \sigma_2,
\end{eqnarray*}
and the pertinent Yukawa couplings are
\begin{eqnarray*}
\lefteqn{
\psi_z^{\overline{u}} \psi_{\overline{z}}^u 
\partial_{\overline{u}} \overline{E}^{\overline{u}} \: + \:
\psi_z^{\overline{v}} \psi_{\overline{z}}^u 
\partial_{\overline{v}} \overline{E}^{\overline{u}} \: + \:
\psi_z^{\overline{u}} \psi_{\overline{z}}^v 
\partial_{\overline{u}} \overline{E}^{\overline{v}} \: + \:
\psi_z^{\overline{v}} \psi_{\overline{z}}^v 
\partial_{\overline{v}} \overline{E}^{\overline{v}}
} \\
& &  \: + \:
\psi_z^{\overline{s}} \psi_{\overline{z}}^s 
\partial_{\overline{s}} \overline{E}^{\overline{s}} \: + \:
\psi_z^{\overline{t}} \psi_{\overline{z}}^t 
\partial_{\overline{t}} \overline{E}^{\overline{t}} \: + \:
\psi_z^{\overline{s}} \psi_{\overline{z}}^t 
\partial_{\overline{s}} \overline{E}^{\overline{t}} \: + \:
\psi_z^{\overline{u}} \psi_{\overline{z}}^t 
\partial_{\overline{u}} \overline{E}^{\overline{t}} \: + \:
\psi_z^{\overline{v}} \psi_{\overline{z}}^t 
\partial_{\overline{v}} \overline{E}^{\overline{t}} 
\end{eqnarray*}
(and their complex conjugates).
In particular, the last three terms,
\begin{displaymath}
\psi_z^{\overline{s}} \psi_{\overline{z}}^t 
\partial_{\overline{s}} \overline{E}^{\overline{t}} \: + \:
\psi_z^{\overline{u}} \psi_{\overline{z}}^t 
\partial_{\overline{u}} \overline{E}^{\overline{t}} \: + \:
\psi_z^{\overline{v}} \psi_{\overline{z}}^t 
\partial_{\overline{v}} \overline{E}^{\overline{t}},
\end{displaymath}
completely encode the nonlinear terms $sf_i(u,v)$ in $E$ -- those nonlinear
terms do not enter into any of the other Yukawa couplings above.

Because the couplings containing the nonlinear terms are not paired
symmetrically, because $\partial_t E^{u,v,s} = 0$,
those nonlinear terms will not affect the
four-fermi contribution to correlation functions.
To see this, first note that when we integrate
over $\psi_z$, $\psi_{\overline{z}}$ zero modes, we will take a
determinant of Yukawa couplings (or rather, what those couplings induce
over the instanton moduli space).  Since determinants are antisymmetric,
and $\partial_t E^{u,v} = 0$, it must be the case that
terms involving $\partial_{u,v} E^t$ can not contribute.
(At a more elementary level, this is saying that since we can
evaluate determinants along either rows or columns, if we choose
to evaluate along a column with only one nonzero entry, then nonzero entries
in the transpose row can not contribute.)

On any compact toric variety, the same will be true more generally.
The point is that for the argument to fail, we need for it to be
possible to build gauge-invariant combinations of the homogeneous
coordinates.  However, that can be done if and only if the toric
variety is noncompact (in which case, the vev of such gauge-invariant
combinations corresponds to noncompact directions).

Let us work through the details in a particular case.
Consider the sector of maps of degree $\vec{d} = (1,-n)$,
which corresponds to maps mapping into the exceptional curve
$E$ with degree $1$.  This case was discussed earlier in this
paper in detail; in effect, we are merely giving a more detailed
derivation of a result used there.
In this case, it is straightforward to compute
\begin{displaymath}
0 \: \longrightarrow \: {\cal F}^* \: \longrightarrow \:
{\cal O}(-1)^4 \oplus {\cal O} \: \stackrel{E'}{\longrightarrow} \:
W \otimes {\cal O} \: \longrightarrow \: 0
\end{displaymath}
over ${\cal M} \cong {\bf P}^3$, with ${\cal F}_1 \cong {\cal O}(-n)^{n-1}$.
In this case, there are only $\psi_{\overline{z}}^{\mu}$ zero
modes for $\mu=s$, hence the only pertinent Yukawa coupling is the
induced coupling\footnote{
In general, we must compute what each coupling $\partial_i E^j$
induces over the moduli space.  In the present case, we need only
consider $\partial_s E^s$, which is independent of $u$, $v$, $s$, and
$t$, and hence takes the same form as it does classically.
}
\begin{displaymath}
\sum_{i} \psi_{z,i}^{\overline{s}} \psi_{\overline{z},i}^s
\left( \gamma_1^* \overline{\sigma}_1 \: + \: \gamma_2^* \overline{\sigma}_2
\right).
\end{displaymath}
Integrating over the $\psi_z$, $\psi_{\overline{z}}$ zero modes gives
a factor of
\begin{displaymath}
\left( \gamma_1^* \overline{\sigma}_1 \: + \: \gamma_2^* \overline{\sigma}_2
\right)^{n-1}.
\end{displaymath}

\end{document}